\documentclass{ar-1col}
\usepackage[numbers]{natbib}
\usepackage{url}
\usepackage{amsfonts,amssymb}
\jname{Xxxx. Xxx. Xxx. Xxx.}
\jvol{AA}
\jyear{YYYY}
\doi{10.1146/((please add article doi))}

\newcommand{\SU}[1]{\ensuremath{\mathrm{SU}( #1 )}}

\newcommand{\SO}[1]{\ensuremath{\mathrm{SO}( #1 )}}

\newcommand{\SpR}[1]{\ensuremath{\mathrm{Sp}( #1,\mathbb{R} )}}
\newcommand{\ket}[1]{\ensuremath{\left| #1 \right\rangle}}
\newcommand{\bra}[1]{\ensuremath{\left\langle #1 \right|}}

\newcommand{\WignerSIXj}[6]
	{
	\left\{
		\begin{array}{ccc}
			#1 & #2 & #3 \\
   			#4 & #5 & #6
		\end{array}
	\right\}
	}
\newcommand{\hw}{\ensuremath{\hbar\Omega}}
\newcommand{\ph}[1]{\ensuremath{#1}p-\ensuremath{#1}h}

\begin{document}

\markboth{Launey et al.}{Nuclear Dynamics and Reactions in the SA Framework}

\title{Nuclear Dynamics and Reactions in the \textit{Ab Initio} Symmetry-Adapted Framework}

\author{Kristina D. Launey,$^{1*}$ Alexis Mercenne,$^{1,2}$ and Tomas Dytrych$^{1,3}$   
\affil{$^1$Department of Physics and Astronomy, Louisiana State University, Baton Rouge, LA 70803, USA
}
\affil{$^2$Center for Theoretical Physics, Sloane Physics Laboratory, Yale University, New Haven, CT 06520, USA}
\affil{$^3$Nuclear Physics Institute of the Czech Academy of Sciences, 250 68 \v{R}e\v{z}, Czech Republic}
\affil{*email: klauney@lsu.edu}
\affil{ORCID Numbers:  0000-0002-8323-7104 (Launey); 0000-0002-2624-3911 (Mercenne); 0000-0002-1554-1462 (Dytrych)}
}

\begin{abstract}
We review the {\it ab initio}  symmetry-adapted (SA) framework for determining the structure of stable and unstable nuclei, along with related electroweak, decay and reaction processes. This framework utilizes the dominant symmetry of nuclear dynamics, the shape-related symplectic \SpR{3} symmetry, which has been shown to emerge from first principles and to expose dominant degrees of freedom that are collective in nature, even in the lightest species or seemingly spherical states.
This feature is illustrated for a broad scope of nuclei ranging from helium to titanium isotopes, enabled by recent developments of  the {\it ab initio} symmetry-adapted no-core shell model expanded to the continuum through the use of the SA basis and that of the resonating group method. The  review focuses on energies, electromagnetic transitions, quadrupole and magnetic moments, radii, form factors, and response function moments, for ground-state rotational bands and giant resonances. The method also determines the structure of reaction fragments that is used to calculate decay widths and alpha-capture reactions for simulated x-ray burst abundance patterns, as well as nucleon-nucleus interactions for cross sections and other reaction observables.

\end{abstract}

\begin{keywords}
ab initio symmetry-adapted framework, 
nuclear structure and reactions, 
nuclear shapes and deformation, 
decay widths and alpha capture reactions, 
x-ray burst abundances, 
nucleon-nucleus potentials
\end{keywords}
\maketitle

\tableofcontents

%
%

\section{INTRODUCTION}

\begin{figure}[h]
\includegraphics[width=\textwidth]{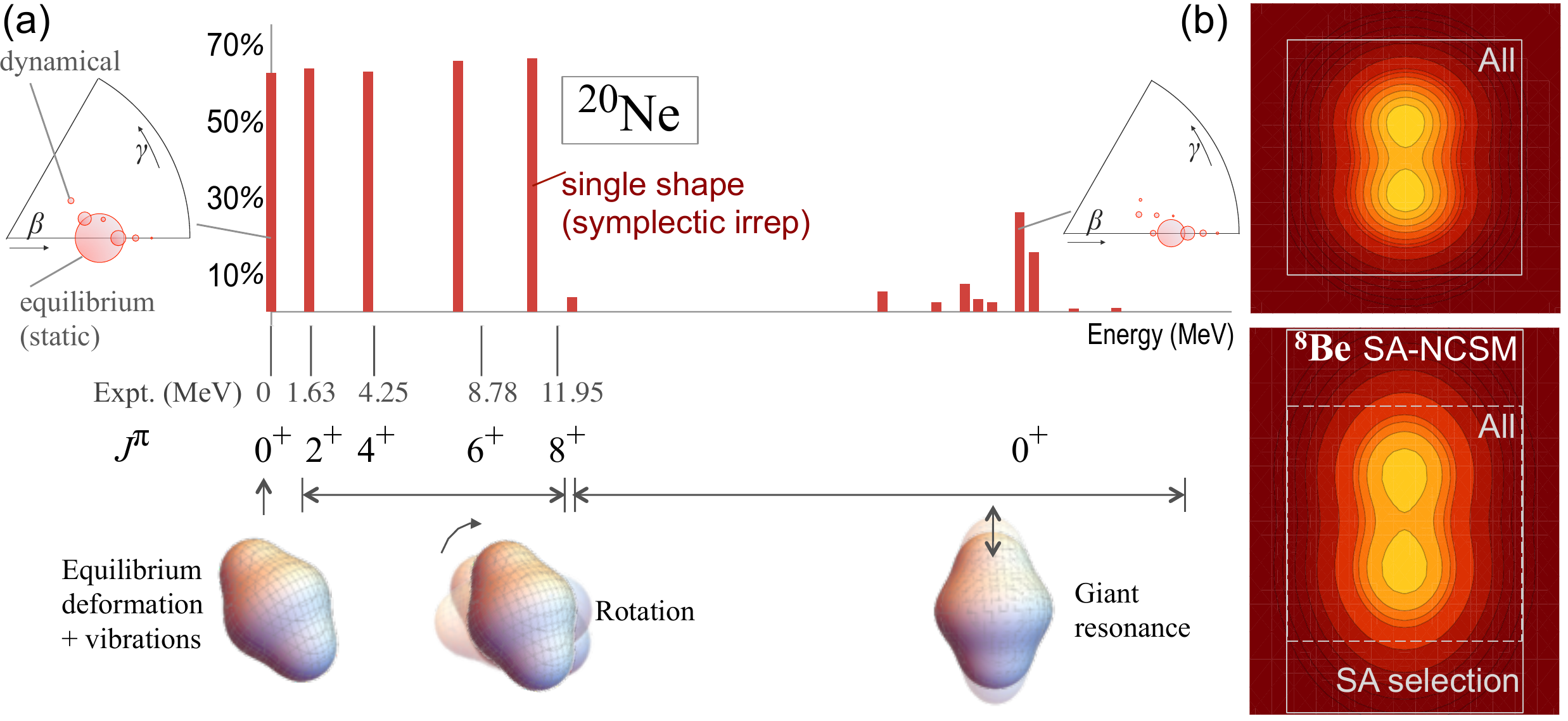}
\caption{(a) Contribution of the most dominant shape to the $0^+$ ground state of $^{20}$Ne and its rotational band ($2^+$, $4^+$, $6^+$, and $8^+$), as well as to excited $0^+$ states, pointing to a fragmented giant monopole resonance; for selected states, the  deformation distribution within a shape is shown in terms of the shape parameters, the average deformation $\beta$ and triaxiality angle $\gamma$  (based on {\it ab initio} SA-NCSM calculations  with NNLO$_{\rm opt}$  in a model space of 11 HO shells with \hw=15 MeV inter-shell distance). 
Figure adapted with permission from Dytrych et al. (2020); copyright 2020 \cite{DytrychLDRWRBB20}.
(b) Schematic illustration of  the SA concept shown for $^8$Be: a smaller model space (square) includes all possible shapes (labeled as ``All") 
and yields spatially compressed wave functions (top); a larger model space (rectangle in lower panel) accommodates, in a well prescribed way, spatially extended  modes (``SA selection") that are  neglected in smaller model spaces. 
}
\label{Sp_pict}
\end{figure}

A fundamental new feature of atomic nuclei has been recently established and shown to naturally emerge from first principles \cite{DytrychLDRWRBB20}. Namely, \textit{ab initio} large-scale calculations have revealed remarkably ubiquitous and only slightly broken symmetry, the \SpR{3} symplectic symmetry, in nuclei up through the calcium region [anticipated to hold even stronger in heavy nuclei \cite{Rowe85}]. Since this symmetry does not mix nuclear shapes, the novel nuclear feature provides important insight from first principles into the physics of  nuclei and their low-lying excitations as dominated by only a few collective shapes -- equilibrium shapes with their vibrations -- that rotate (Fig. \ref{Sp_pict}a).  

This remarkable outcome builds upon a decades-long research, starting with the pivotal work of Draayer \cite{DraayerSU3_1,DraayerWR84,BlokhinBD95,LauneyDD16} and that of Rowe and Rosensteel \cite{RosensteelR77,RosensteelR80,Rowe85,Rowe96}, who have successfully harnessed group theory as a powerful tool for understanding and computing the intricate structure of nuclei. This pioneering work  has been instrumental in designing the theory that underpins many highly ordered patterns unveiled amidst the large body of experimental data \cite{HeydeW11,Wood16,RoweW18}, while explaining phenomena observed in energy spectra, $E2$ transitions and deformation, giant resonances (GR), scissor modes and $M1$ transitions, electron scattering form factors, as well as the interplay of pairing with collectivity. The new developments and insights have provided the critical structure raised upon the very foundation laid by Elliott \cite{Elliott58,Elliott58b,ElliottH62} and Hecht \cite{HechtA69,HechtZ79}, and opened the path for large-scale calculations  feasible today on  supercomputers. Now, within  an {\it ab initio} framework without {\it a priori} symmetry  assumptions, the symmetry-adapted no-core shell model (SA-NCSM) \cite{DytrychSBDV_PRL07,DytrychLMCDVL_PRL12,LauneyDD16} with chiral effective field theory (EFT) interactions \cite{BedaqueVKolck02,EpelbaumNGKMW02,EntemM03}, not only explains but also predicts the emergence of nuclear collectivity across nuclei, even in close-to-spherical nuclear states without any recognizable rotational properties.
\begin{marginnote}[]
\entry{SA-NCSM}{symmetry-adapted no-core shell model}
\entry{SA}{symmetry-adapted }
\entry{irrep}{irreducible representation}
\entry{EFT}{effective field theory}
\entry{GR}{giant resonance}
\end{marginnote}

The symmetry-adapted (SA)  framework \cite{DytrychSBDV_PRL07,LauneyDD16,DytrychLDRWRBB20}, discussed in Sec. \ref{SA}, capitalizes on these findings and presents solutions  in terms of a physically relevant basis of nuclear shapes. By exploiting this approximate symmetry, the SA framework resolves the scale explosion problem in nuclear structure calculations, {\em i.e.}, the explosive growth in computational resource demands with increasing number of particles and size of the spaces  in which they reside (referred to as ``model spaces''). It is based on the idea that the infinite Hilbert space can be equivalently spanned by ``microscopic" nuclear shapes and their rotations [or symplectic  irreducible representations (irreps), subspaces that preserve the symmetry], where  ``microscopic" refers to the fact that these configurations track with position and momentum coordinates of each particle. A collective nuclear shape can be viewed as  an equilibrium (``static") deformation and its vibrations (``dynamical" deformations) of the GR type (cf. Sec. \ref{sancsm}), as illustrated in the $\beta$-$\gamma$ plots of Fig. \ref{Sp_pict}a~\cite{Rowe13,DytrychLDRWRBB20}. A key ingredient of the SA concept is illustrated in Fig. \ref{Sp_pict}b, namely,  while many  shapes relevant to low-lying states are included in typical shell-model spaces,  the vibrations of largely deformed equilibrium shapes and  spatially extended modes like clustering often lie outside such spaces. The selected model space in the SA framework remedies this, and includes, in a well prescribed way, those  configurations. Note that this is critical for enhanced deformation, since spherical and less deformed shapes easily develop in comparatively small model-space sizes. 
\begin{marginnote}[]
\entry{``Static" deformation}{Equilibrium shape, invariant under \SpR{3} transformations}
\entry{``Dynamical" deformation}{GR-type vibration of an equilibrium shape}
\entry{Collective nuclear shape}{Equilibrium shape with its vibrations; together with its rotations span a single \SpR{3} irrep}
\end{marginnote}

{\it Ab initio} descriptions of spherical and deformed nuclei up through the calcium region are now possible without the use of interaction renormalization procedures, as discussed in  Sec. \ref{SA}. In particular, Refs. \cite{DytrychLMCDVL_PRL12,DytrychMLDVCLCS11,DytrychHLDMVLO14,LauneyDD16,BakerLBND20} have shown that the SA-NCSM can use significantly reduced  model spaces as compared with the corresponding  ultra-large  conventional model spaces without compromising the accuracy of results for various observables. This allows the SA-NCSM to accommodate larger model spaces  and to reach heavier nuclei, such as $^{20}$Ne \cite{DytrychLDRWRBB20}, $^{21}$Mg \cite{Ruotsalainen19}, $^{22}$Mg \cite{Henderson:2017dqc}, $^{28}$Mg \cite{PhysRevC.100.014322},  as well as $^{32}$Ne and $^{48}$Ti \cite{LauneySOTANCP42018}.

This makes the SA basis especially suitable for describing nuclear reactions, key to understanding processes measured in experiments and those
in extreme environments, from stellar explosions to the interior of nuclear reactors. Remarkable progress has been recently made in first-principles many-body approaches to  scattering and nuclear reactions for light nuclei (for an overview, see \citenum{FRIBTAwhite2018}), 
including studies of elastic scattering \cite{NollettPWCH07,HagenDHP07,PhysRevLett.101.092501,ElhatisariLRE15,PhysRevLett.125.112503}, photoabsorption \cite{PhysRevC.90.064619}, transfer \cite{NavratilQ12} and capture reactions \cite{PhysRevLett.105.232502}, as well as thermonuclear fusion \cite{HupinQN19}. Expanding the reach of \textit{ab initio} reactions beyond the lightest species, including deformed targets -- from helium to calcium isotopes -- as well as alpha projectiles,  is now feasible with the SA basis, and we review three recent developments  in Sec. \ref{SAreactions}.
We start with a remarkable illustration, namely, the first description of the $\alpha+^{16}$O system based on \textit{ab initio} SA-NCSM descriptions of $^{20}$Ne, along with an estimate for the alpha capture  reaction rate $^{16}$O$(\alpha,\gamma)^{20}$Ne at temperatures relevant to x-ray burst (XRB) nucleosynthesis \cite{DreyfussLESBDD20}. 
\begin{marginnote}[]
\entry{XRB}{x-ray burst }
\entry{RGM}{resonating group method}
\entry{SA-RGM}{symmetry-adapted resonating group method}
\end{marginnote}

For a single-nucleon projectile, the SA basis plays a key role in  the recently developed \textit{ab initio} symmetry-adapted resonating group method (SA-RGM) \cite{MercenneLEDP19,Mercenne2LEDP19,Mercenne:2019LDEP} for cross sections of reactions and scattering at low-energy reactions, which is the astrophysically relevant energy regime. It follows  the successful merging  of the resonating-group method (RGM) \cite{WildermuthT77} with the   no-core shell model (NCSM) for light nuclei \cite{QuaglioniN09}, which provided  unified descriptions of structure and reaction observables from first principles. The SA-RGM  utilizes the same symmetry considerations as for the SA-NCSM, and in doing so, it empowers the approach with the capability to simultaneously describe both bound and scattering states, while preserving  the Pauli exclusion principle and translational invariance (see Sec. \ref{sargm} for the n+$^{16}$O and n+$^{20}$Ne systems, with a focus on low-lying resonant and scattering states). 
For intermediate energy,  which corresponds to current experimental studies at rare isotope beam facilities, the spectator expansion of the multiple scattering theory \cite{Elster:1996xh,Dussan:2014vta}  has recently offered a fully consistent \textit{ab initio} approach to nucleon scattering that  accounts for the spin of the struck nucleon in the target \cite{BurrowsBEWLMP20}, as well as  for the microscopic structure of the target from first principles by utilizing \textit{ab initio} one-body nuclear densities \cite{Burrows:2017wqn} (see Sec. \ref{multiscatt} for proton scattering on $^4$He and $^{20}$Ne targets, at projectile laboratory energies of 100-200 MeV per nucleon). As an important outcome
these frameworks
offer a way to construct nucleon-nucleus effective interactions rooted in first principles, the key ingredient in reaction theory (see Sec. \ref{fewbody}).

The overarching goal is -- by exploiting dominant symmetries in nuclear dynamics and the SA basis --  to provide reliable descriptions of nuclear reactions that can be measured at rare isotope beam facilities and that are of particular interest in astrophysics. For example, the proton-capture $^{23}$Al(p,$\gamma)^{24}$Si reaction is one of the several reactions identified to have a substantial effect on luminosity profiles in time (light curves) from XRB nucleosynthesis simulations \cite{Cyburt10,PhysRevLett.122.232701}. Predictions for XRB light curves are important, because they are available from observational astronomy  (see, e.g., \citenum{Galloway_2008}).
Equally important are ($\alpha$,p),  ($\alpha$,n), and ($\alpha$,$\gamma$) reactions~\cite{Bru15,Wiescher_ARAA50_2012}, and especially
the $^{12}$C$(\alpha,\gamma)^{16}$O reaction rate \cite{RevModPhys.89.035007}, one of the most important reaction  to stellar helium burning, that currently has uncertainties that may potentially impact predicted accuracy of the final black hole mass \cite{Farmer_2020,McDermott20} in analysis of current and upcoming gravitational wave interferometer detections of binary black hole mergers \cite{PhysRevLett.119.161101}.  
Furthermore, measuring neutron capture cross sections is critical to  astrophysical simulations that aim to resolve the r-process~\cite{PhysRevC.79.045809}. While direct capture measurements with exotic isotopes are often not possible, due to practical considerations such as very small cross sections, unavailability of beams, or the infeasibility of measuring neutron-induced reactions on radioactive isotopes, the one-nucleon transfer reaction (d,p) has been proposed as suitable indirect tool for providing information about cross sections for neutron capture reactions (see, e.g., \citenum{RevModPhys.84.353,PhysRevLett.121.052501}). 
In addition, the use of n$+{}^{48}{\rm Ca}$ scattering and total neutron cross section measurements can provide constraint on the neutron skin thickness~\cite{mahzoon:2017}, important for pinpointing
the equation of state of neutron-star matter (e.g., see \citenum{PhysRevLett.120.172702}).

\section{NUCLEAR APPROACHES IN THE ERA OF RARE ISOTOPE BEAM FACILITIES}
\label{fewbody}

Currently, only a small fraction of the thousands of nuclei that exist can be measured and described reasonably well by theory. Most of these lie in the ``valley of stability". This underlines the need for exploration of and beyond the drip lines, that is, the limit of nuclear stability with respect to the emission of one nucleon. Measuring and describing nuclei far from stability is indeed of great importance for nuclear astrophysics, as many short-lived nuclei are formed during cataclysmic events in the Universe, and can, in turn, largely influence various astrophysics simulations. 
As measurements involve scattering and reactions of nuclei, it is important to have a reliable and predictive theoretical  framework of reaction processes that is applicable to stable and unstable nuclei. 

\begin{textbox}[h]\section{Rare isotope beam facilities and needs for theory}
Experiments at current and upcoming rare isotope beam facilities can probe nucleon-nucleon interactions and nuclear structure, but require novel theoretical approaches that can reliably model reactions of short-lived isotopes to support and inform experimental programs.  Historically, two major cornerstone frameworks have been developed: (1)  Few-body techniques (with early applications to reactions) use correct asymptotics (i.e., the wave function of the reaction fragments at large distances), but may often neglect the microscopic structure of the clusters  and employ optical potentials fitted to elastic scattering data of stable nuclei (see, e.g., \citenum{Furumoto:2019anr,Weppner:2009qy,koning:2003zz}). (2) Many-body techniques (with early applications to structure) use many-body degrees of freedom and target unified  structure and reaction descriptions, 
but may often neglect or partially account for the continuum and are often limited in mass or number of active particles, as a result of  increased complexity. 
Recent developments have started to address many of these challenges by merging both concepts, by including microscopic degrees of freedom in few-body models,   by constructing microscopic few-body effective interactions (optical potentials) \cite{idini19,RotureauDHNP17,BurrowsBEWLMP20}, as well as by including continuum and collective degrees of freedom into many-body approaches \cite{HupinQN19,MercenneLEDP19,DreyfussLESBDD20,MichelNP02,mercennemp19}.
The new physics to be learned from proposed experiments and new theoretical developments for unstable nuclei  are summarized in several recent experimental and theoretical white papers (\citenum{ARCONES20171,CARLSON201768,FRIBTAwhite2018} and references therein).
\end{textbox}

Exact solutions for the scattering problem are only available for systems with up to five nucleons  \cite{Fonseca:2017koi,Deltuva:2017bia,fad2b,Viviani:2016cww}. 
Nuclear approaches to reactions and scattering face several challenges, especially since nuclear probes are  often peripheral and, hence, require a correct asymptotic treatment. Major challenges include: the long-range Coulomb force, and in particular in the case of large projectile and/or target charges where the asymptotics may not be analytically known;  the high sensitivity of reaction observables  to the reaction thresholds ($Q$-values); the importance of the non-resonant continuum when nuclei break up into the continuum; and the difficulties in describing scattering states asymptotics  with single-particle bound-state bases typically used in many-body methods  \cite{FRIBTAwhite2018}.

Currently, many successful reaction models employ approximations and largely rely on constraints from data (phenomenology), including $R$-matrix methods, Glauber theory, the Hauser-Feshbach model, phenomenological optical potentials, and the  valence shell model. While these methods have been very successful in certain mass regions and energies across the valley of stability, they are often limited by the approximations they assume. For example, the Hauser-Feshbach model assumes high level densities; phenomenological optical potentials do well at comparatively high projectile energy, whereas at low energies they fail to account appropriately for isolated resonances, and in addition, they are fitted to stable nuclei and  uncertainties become uncontrolled as one moves away from stability  \cite{LovellN15}; reaction models often assume no structure of the clusters; and valence shell-model calculations omit particle-hole excitations that are expected to play an important role in weakly bound systems. 

Alternatively, many-nucleon approaches with controlled approximations may be employed (see, e.g., the recent reviews \citenum{FRIBTAwhite2018,Hergert20}).
These include using a physically relevant basis, such as the symmetry-adapted basis, that accommodates large enough model spaces necessary to describe the wave function tail within the potential  effective range, while  at large distances the exact Coulomb wave functions are used  \cite{DreyfussLESBDD20} (see Sec. \ref{alpha}); adding  
a basis that explicitly considers the reaction fragments, such as the RGM basis \cite{PhysRevLett.101.092501,QuaglioniN09,HupinQN19,MercenneLEDP19} (see Sec. \ref{sargm}); and starting with a complex-momentum single-particle basis, such as the  Berggren basis (e.g., see \citenum{MichelNP02,RotureauMNPD06,FossezRMP17,mercenne16,mercennemp19}), which imposes single-particle scattering boundary conditions, thereby consistently treats bound states, resonances and scattering states within the same framework.
However, these methods are often limited by computational resources and may not achieve the required level of accuracy.  In such cases, it might be advantageous to adopt a hybrid approach that allows some quantities to be directly taken from (or strongly constrained by) data, such as threshold measurements.
Thus, for example, experiments can provide precise thresholds, whereas theory can pinpoint critical collective and clustering correlations in wave functions to achieve the best estimates for reaction rates for astrophysics.
Indeed, to analyze and interpret experimental data, theory with uncertainties lower than 10\% is needed \cite{FRIBTAwhite2018}.

\section{SYMMETRY-ADAPTED FRAMEWORK AND ROLE OF SYMPLECTIC SYMMETRY}
\label{SA}

\subsection{Symmetry-Adapted No-Core Shell Model}
\label{sancsm}

{\it Ab initio} approaches build upon a ``first principles" foundation, namely, the properties of only two or three nucleons that are often tied to symmetries and symmetry-breaking patterns of the underlying quantum chromodynamics theory.
We utilize the {\it ab initio} nuclear shell-model theory \cite{NavratilVB00,BarrettNV13} that solves the many-body Schr\"odinger equation for $A$ particles,
\begin{equation}
H \Psi(\vec r_1, \vec r_2, \ldots, \vec r_A) = E \Psi(\vec r_1, \vec r_2, \ldots, \vec r_A), {\rm with}\, H = T_{\rm rel} + V_{NN}  + V_{3N} + \ldots + V_{\rm Coulomb}.
\label{ShrEqn}
\end{equation}
In its most general form, it is an exact many-body ``configuration interaction" method, for which the interaction and basis configurations are as follows. 
The intrinsic non-relativistic nuclear  Hamiltonian $H$ includes
the relative kinetic energy $T_{\rm rel} =\frac{1}{A}\sum_{i<j}^A\frac{(\vec p_i - \vec p_j)^2}{2m}$ ($m$ is the nucleon mass),  
the  nucleon-nucleon (NN)  and, possibly, three-nucleon (3N)  interactions, typically derived in the chiral effective field theory \cite{BedaqueVKolck02,EpelbaumNGKMW02,EntemM03},
along with the Coulomb interaction between the protons. 
A complete orthonormal many-particle basis $\psi_k$ is adopted, e.g.,  the antisymmetrized products of  single-particle states  of a spherical harmonic oscillator (HO) of  characteristic length $b=\sqrt{\hbar \over m\Omega}$ and frequency $\Omega$. 
The expansion $\Psi(\vec r_1, \vec r_2, \ldots, \vec r_A) = \sum_{k} c_k \psi_k(\vec r_1, \vec r_2, \ldots, \vec r_A)$
 renders Eq. (\ref{ShrEqn}) into a matrix eigenvalue equation,
$
\sum_{k'} H_{k k'} c_{k'} = E c_k,
$
with unknowns $c_k$, where the many-particle Hamiltonian matrix elements
$H_{k k'} = \langle \psi_k | H | \psi_{k'} \rangle$ are calculated for the given interaction and the solution $\{c_k^2\}$ defines a set of probability amplitudes. 
\begin{marginnote}[]
\entry{HO}{harmonic oscillator}
\entry{NN}{nucleon-nucleon}
\entry{3N}{three-nucleon}
\end{marginnote}

We note that, throughout this paper, we adopt the term ``\textit{ab initio}" for a system of $A$ particles in cases when an $A$-body approach with controlled approximations are employed, such as the SA-NCSM, together with realistic interactions that reproduce NN phase-shift data to a given energy with high precision (and perhaps properties of three-nucleon systems), such as JISP16 \cite{ShirokovMZVW07}, AV18 \cite{WiringaSS95}, along with chiral potentials, N3LO-EM \cite{EntemM03}, NNLO$_{\rm opt}$ \cite{Ekstrom13}, and NNLO$_{\rm sat}$ \cite{PhysRevC.91.051301}, including the complementary 3N forces. 

An important feature of the symmetry-adapted framework is that the model space is reorganized to a symmetry-adapted basis that respects the deformation-related \SU{3} symmetry or the shape-related \SpR{3} symmetry \cite{LauneyDD16}. We note that while the model utilizes symmetry groups to construct the basis, calculations are not limited {\it a priori} by any symmetry and employ a large set of basis states that can,  if the nuclear Hamiltonian demands, describe a significant symmetry breaking.
The SA-NCSM is reviewed in Ref. \cite{LauneyDD16} and has been first applied to light nuclei using the \SU{3}-adapted basis \cite{DytrychLMCDVL_PRL12} and soon expanded with an \SpR{3}-adapted basis and to heavier nuclei \cite{LauneyDD16,DytrychLDRWRBB20}. Both bases are briefly discussed next.
\begin{textbox}[H]\section{Deformation-related \SU{3} and shape-related symplectic \SpR{3}  groups}
A nuclear shape is microscopically described by a set of $A$-particle configurations that preserves the \SpR{3} symmetry and includes an equilibrium deformation and its vibrations, the dynamical deformations, along with rotations \cite{Rowe13,DytrychLDRWRBB20}. From a mathematical point of view, the symplectic group \SpR{3} consists of all {\it particle-independent} linear canonical transformations of the single-particle phase-space observables, the positions $\vec r_i$ and momenta $\vec p_i$,  that preserve the Heisenberg commutation relations $[r_{i\alpha},p_{j\beta}]=i\hbar \delta_{ij}\delta_{\alpha \beta}$ (with particle index $i=1,\dots, A$ and spacial directions $\alpha,\beta=x,y,z$) \cite{Rowe85,Rowe13,LauneyDD16}. 
A key feature is that several physically relevant operators do not mix nuclear shapes, including the total kinetic energy, $\frac{p^2}{2}=\frac{1}{2}\sum_{i=1}^A{\vec p_i \cdot \vec p_i}$, the monopole moment, $r^2=\sum_{i=1}^A{\vec r_i \cdot \vec r_i}$, the quadrupole moment, $Q_{2M}=\sqrt{16\pi/5 }\sum_{i=1}^A r_i^2Y_{2M}(\hat r_i)$, the orbital momentum, $\vec L=\sum_{i=1}^A \vec r_i \times \vec p_i$, and the many-body harmonic oscillator Hamiltonian, $H_0=\frac{p^2}{2}+\frac{r^2}{2}$. A subset of these act only within a single deformation, or an \SU{3} irrep, namely, the operators $Q_{2M}$, when restricted to a single shell, and $L$.
\end{textbox}

\noindent
{\bf \SU{3}-adapted basis.} The  many-nucleon basis states of the SA-NCSM are constructed using efficient group-theoretical algorithms and are labeled according to  \SU{3}$_{(\lambda\,\mu)}\times$\SU{2}$_S$ by the total  intrinsic spin $S$ and $(\lambda\,\mu)$ quantum numbers with $\lambda=N_z-N_x$ and $\mu=N_x-N_y$, where $N_x+N_y+N_z=N_0+N$, for a total of $N_0+N$ HO quanta distributed in the $x$, $y$, and $z$ direction. Here, $N_0\hw$ is the lowest total HO energy for all particles (``valence-shell configuration") and $N\hw$ ($N\le N_{\rm max}$) is the additional energy of all particle-hole excitations. Hence, $N_x=N_y=N_z$, or equivalently $(\lambda\,\mu)=(0\, 0)$, describes a spherical configuration, while $N_z$ larger than  $N_x=N_y$ ($\mu=0$) indicates prolate deformation. Clearly, a closed-shell configuration has $(0\, 0)$, and
spherical shapes, or no deformation, are a part of the SA basis. However, most nuclei --  from light to heavy -- are deformed in the \emph{body-fixed} frame ($N_z > N_x > N_y$), which  for $0^+$ states,  appear  spherical in the \emph{laboratory} frame.

\noindent
{\bf \SpR{3}-adapted basis.} Furthermore, considering the embedding \SpR{3} symmetry according to \SpR{3}$\supset $\SU{3}, one can further organize \SU{3}  deformed configurations into symplectic  irreps, subspaces that preserve the \SpR{3}  symmetry. A symplectic irrep is characterized by  a given equilibrium shape, labeled by a single deformation $N(\lambda\,\mu)$. For example, a symplectic irrep $0(8\,0)$ in $^{20}$Ne consists of a prolate $0(8\,0)$ equilibrium shape with $\lambda=8$ and $\mu=0$ in the valence-shell \ph{0} (0-particle-0-hole) subspace, along with many other \SU{3} deformed configurations (vibrations), such as $2(10\, 0)$, $2(6\,2)$ and $8(16\,0)$, that include particle-hole excitations of the  equilibrium shape to higher shells (for further details, see \citenum{LauneyDD16,DytrychLDRWRBB20,LauneyDSBD20}). 
These vibrations are multiples of 2\hw~\ph{1} excitations of the giant-resonance monopole and quadrupole types, that is, induced by the monopole $r^2$ and quadrupole $Q$ operators, respectively.

A major advantage of the SA-NCSM is that  the SA model space can be down-selected to a subset of SA basis states that describe equilibrium and dynamical deformation, and within this selected model space the spurious center-of-mass motion can be factored out exactly~\cite{Verhaar60,Hecht71}.  Another major advantage is that deformation and collectivity is properly treated in the approach \emph{without} the need for breaking and restoring rotational symmetry. The reason is that basis states utilize the \SU{3}$\supset$ \SO{3} reduction chain that has a good orbital momentum, whereas all \SU{3} reduced matrix elements depend only on $(\lambda\,\mu)$ and can be calculated in the simpler canonical \SU{3}$\supset$ \SU{2}  reduction chain that takes advantage of the Cartesian scheme ($N_z,N_x,N_y$).
A third major advantage is the use of group theory, including the Wigner-Eckart theorem and group-theoretical algorithms (e.g., see \citenum{DraayerLPL89,LangrDLD18,1937-1632_2019_0_183}).

\subsection{Unveiling Dominant Features and Symmetries: Equilibrium Shapes, Vibrations, and Rotations }

\begin{figure}[h]
\includegraphics[width=1\textwidth]{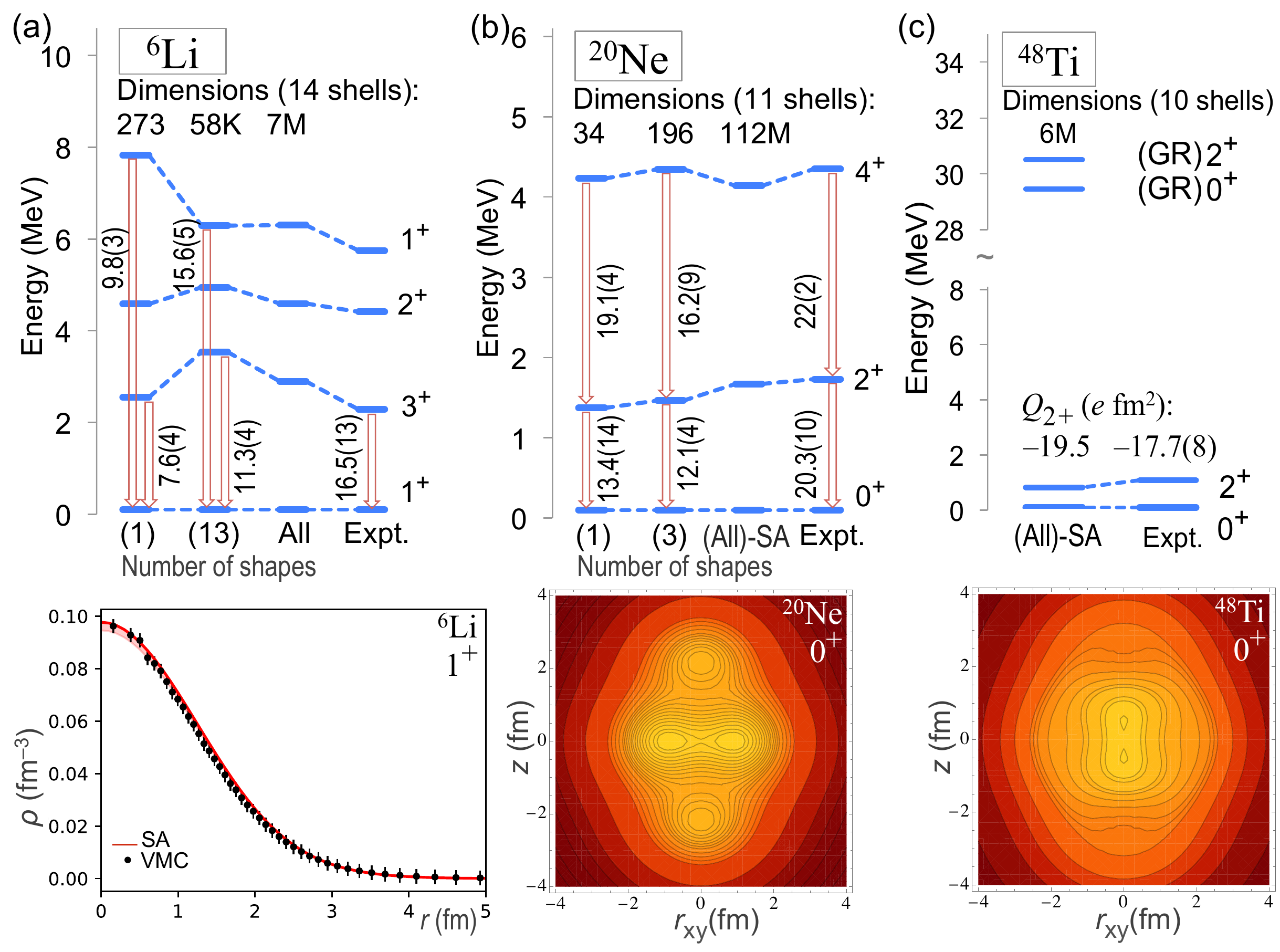}
\caption{Energy spectra, $B(E2)$ transition strengths (in W.u.) or $Q$ quadrupole moment (in $e$ fm$^2$), and one-body densities for (a) $^{6}$Li, (b) $^{20}$Ne, and (c) $^{48}$Ti, based on {\it ab initio} SA-NCSM calculations  with NNLO$_{\rm opt}$. Energy spectra are labeled by the dimensions of the largest model spaces used and the number of \SpR{3} irreps (shapes). Observables are reported for infinite model spaces, except calculations with all symplectic irreps (``All")  in complete or SA spaces that use \hw=15 MeV. Point-proton densities are shown for the largest model spaces [red solid line in (a) and spatial profiles in (b) \& (c)], as well as for the single irrep [red band in (a)].
The $^{6}$Li  point-proton density is compared with VMC with AV18/Urbana IX \cite{PhysRevC.84.024319}.
Part of figure adapted with permission from Dytrych et al. (2020); copyright 2020 \cite{DytrychLDRWRBB20}.}
\label{enspectra}
\end{figure}

As mentioned above, a remarkable outcome has been recently reported, as unveiled from first-principle SA-NCSM calculations below the calcium region, that nuclei exhibit relatively simple physics \cite{DytrychLDRWRBB20}. We now understand that a low-lying nuclear state
is predominantly composed  of a few equilibrium shapes that vibrate through excitations of the giant-resonance monopole and quadrupole type, and rotate as well (see also \citenum{Johnson15,Johnson20}). 
Specifically, nuclei are predominantly comprised -- typically in excess of 70-80\% -- of only  a few 
 shapes, often a single shape (a single symplectic irrep) as for, e.g.,  the odd-odd $^{6}$Li  (Fig. \ref{enspectra}a) and $^{8}$B, the cluster-like $^{8}$Be (Fig. \ref{Sp_pict}b), $^{16}$O (often considered to be closed-shell), and the intermediate-mass $^{20}$Ne (Figs. \ref{Sp_pict}a \& \ref{enspectra}b), or two shapes, e.g., for $^{8}$He (generally considered to be spherical) and $^{12}$C \cite{DytrychMLDVCLCS11} [see also results in \citenum{LauneyDD16,Johnson20} based on \SU{3} analysis]. Hence, the ground state of $^{6}$Li and $^{20}$Ne ($^{16}$O) is found to have a dominant prolate (spherical) shape,  while an oblate shape dominates in the cases of $^{8}$He and $^{12}$C.  The symplectic symmetry has been found to hold even in excited states, as shown in Ref. \cite{LauneyDSBD20}, and for $^{7}$Be \cite{PhysRevLett.125.102505}.

Besides the predominant irrep(s), there is a manageable number of symplectic irreps, each of which contributes at a level that is typically at least an order of magnitude  smaller, as discussed in  Ref.~\cite{DytrychLDRWRBB20}. In addition, this study has shown that realistic interactions yield practically the same symplectic content in low-lying states as the one observed in the ground state (see Fig. \ref{Sp_pict}a for $2^+$, $4^+$, $6^+$, and $8^+$), which is a rigorous signature of rotations of a shape and can be used to identify members of a rotational band. 

By exploiting the approximate symplectic symmetry, excitation energies and $B(E2)$ transition strengths are studied for selected nuclei using only a few symplectic irreps or  \SU{3} model spaces (which include all symplectic irreps),  as shown in Fig. \ref{enspectra} for $^6$Li and $^{20}$Ne. Within a few symplectic irreps, these observables show a relative fast convergence trend across variations in the model space size and resolution (related to $N_{\rm max}$ and \hw) \cite{DytrychLDRWRBB20,LauneyDSBD20}, yielding extrapolations to infinitely many  shells  with typical errors of $\sim 100$ keV for excitation energies  and  of $\sim 4\%$ for $B(E2)$. We note that $E2$ transitions are determined by the quadrupole operator $Q$, an  \SpR{3} generator that does not mix symplectic irreps -- the predominance of a single symplectic irrep reveals the remarkable result that the largest fraction of these transitions, 
and hence nuclear collectivity,  necessarily emerges within this symplectic irrep [similarly for rms radii, since $r^2$ is also an  \SpR{3} generator]. 

We note the small  model-space size used for computations of low-lying states in $^6$Li and $^{20}$Ne (listed under ``Dimensions'' in Fig. \ref{enspectra}). For comparison, the corresponding NCSM dimension for $J^\pi=0^+,2^+,4^+$ in $^{20}$Ne in 11 HO shells  is $3.8\times 10^{10}$. It is then remarkable that even excitation energies calculated in model spaces selected down to a few symplectic irreps closely reproduce the experimental data.

\subsection{Benchmark Studies and Nuclear Properties}
\label{benchmarkcalc}

This section summarizes the results of a series of benchmark studies, in which the SA-NCSM has been shown to use significantly reduced selected model spaces as compared with the corresponding  large 
complete $N_{\rm max}$  model space (or equivalently, NCSM) without compromising the accuracy for various observables that probe nuclear properties. These include energies, point-particle 
rms radii, electric quadrupole and magnetic dipole moments, reduced  $B(E2)$ transition strengths \cite{DytrychLMCDVL_PRL12,DytrychMLDVCLCS11},  electron scattering form factors \cite{DytrychHLDMVLO14}, and sum rules \cite{BakerLBND20}. Indeed, results for light nuclei (with the illustrative examples of $^4$He, $^6$Li, and $^{12}$C presented below) agree with those of other \textit{ab initio} approaches, such as the hyperspherical harmonics (HH),  no-core shell model, as well as  variational  (VMC) and Green's function (GFMC) Monte Carlo. 
Following this, we illustrate the capability of the the SA concept   to reach
heavier nuclei, such as $^{32}$Ne and $^{48}$Ti \cite{LauneySOTANCP42018}.
\begin{figure}[h]
\includegraphics[width=\textwidth]{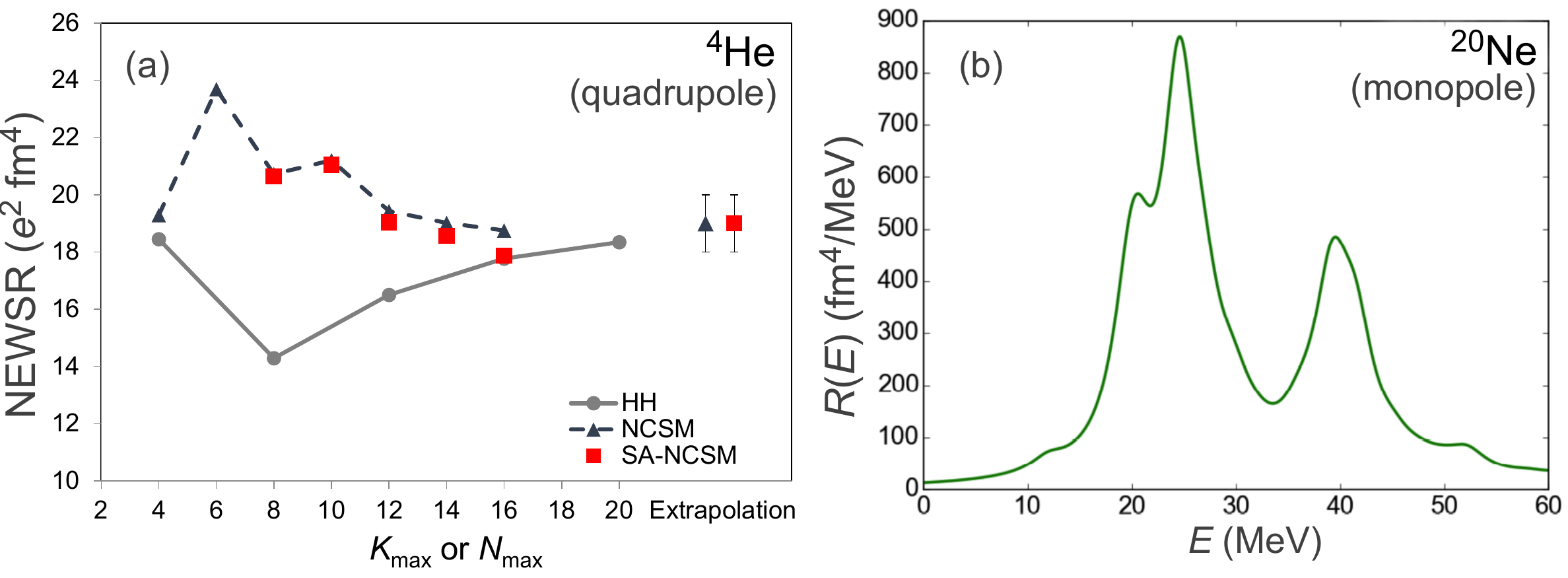}
\caption{(a) Quadrupole NEWSR (translationally invariant) for $^4$He as a function of the model-space size ($K_{\mathrm{max}}$ or $N_{\mathrm{max}}$), based on {\it ab initio} HH and SA-NCSM ($\hw=25$ MeV)  calculations  with N3LO-EM interaction (NN only).  
Figure adapted with permission from Baker et al. (2020); copyright 2020 \cite{BakerLBND20}.
(b) Lorentz integral transform (translationally invariant) for monopole transitions to the $^{20}$Ne ground state vs. excitation energy $E$ for a Lorentzian kernel width of $\Gamma_{\mathfrak{L}}=2$ MeV that yields the  response function  in the $\Gamma_{\mathfrak{L}} \rightarrow 0$ limit \cite{BakerLBND20} 
(based on {\it ab initio} SA-NCSM calculations  with NNLO$_{\rm opt}$  and  \hw=15 MeV). Figure adapted with permission from Launey et al. (2020); copyright 2020  \cite{LauneyDSBD20}.
}
\label{ff_lsr}
\end{figure}

In particular, for $^4$He, Ref. \cite{BakerLBND20} has compared SA-NCSM observables with exact solutions of the HH. 
We present selected results of this benchmark study, with a focus  on  the ground-state (g.s.) energy and point-proton rms radius of $^4$He with the JISP16 and N3LO-EM potentials, as well as selected energy moments of the response function, or so-called sum rules. Response functions  for electromagnetic probes are important, because they are used to calculate cross sections and can reveal information about the dynamical structure of the nucleus itself. While it is desirable to compute the full response function, it is sometimes easier to study its energy moments, which can be compared with experiment as well.
The SA-NCSM calculations, when  extrapolated to infinite spaces, are found to practically coincide with the HH and NCSM results (Table \ref{benchmark}), while  exhibiting very good convergence with the model-space size, parameterized by $K_{\rm max}$ for the HH and  $N_{\rm max}$ for the SA-NCSM and the NCSM (see Fig. \ref{ff_lsr}a for an illustrative example). Overall,  sum rules,  such as the non-energy weighted sum rule (NEWSR),  energy-weighted sum rule (EWSR), and inverse energy-weighted sum rule (IEWSR), for monopole, dipole, and quadrupole probes show agreement within $2\sigma$  between the HH results and the extrapolated SA-NCSM values for JISP16 (see Table \ref{benchmark} for selected sum rules and interactions, whereas a complete set of values for JISP16, N3LO, and NNLO$_{\rm opt}$ is available in Tables II and III of \citenum{BakerLBND20}). 
In Table \ref{benchmark}, extrapolated values for NCSM and SA-NCSM are based on several model-space sizes  up to 17 shells and a 10\% variation in the $\hw$ parameter; the HH results without uncertainties are reported at convergence. We note that all observables reported are translationally invariant, which is not trivial for sum rules calculated in many-body methods that use laboratory-frame coordinates and has been resolved in Ref. \cite{BakerLBND20} by a novel algorithm based on the Lawson procedure \cite{Lawson74}.
\begin{marginnote}[]
\entry{NEWSR}{non-energy weighted sum rule}
\entry{EWSR}{energy weighted sum rule}
\entry{IEWSR}{inverse energy weighted sum rule}
\end{marginnote}
\begin{table}[h]
\tabcolsep7.5pt
\caption{
Benchmark results for the SA framework compared with other \textit{ab initio} methods for selected observables (see text for details). Unless otherwise stated, calculations with the JISP16 potential and experimental data from Ref. \cite{Tilley2002}.
}
\label{benchmark}
\begin{center}
\begin{tabular}{@{}l|l|l|l|l@{}}
\hline
		&	 SA-NCSM 	&		 & & 			Experiment	\\
\hline
\multicolumn{1}{c|}{ {\bf $^4$He $0^+_{\rm g.s.}$}  } 	 &		&		{\bf NCSM}$^{\rm a}$	&				{\bf HH}$^{\rm a}$	&	\\	
		BE (MeV)	&	28.2944(7)		&	28.2986(4)		&	28.300		&	28.30	\\
			&	26(1)$^*$ 	&	25(1)	$^*$ &	$25.3(1)^*$		&		\\
		 Dipole NEWSR ($e^2$ fm$^2$)	&	0.94(2)$^*$	&	0.95(3)$^*$	&	0.945$^*$		&	N/A	\\
		 Monopole EWSR/NEWSR (MeV) 	&	 6.67(3)		&	 6.63(1)		&	 6.623(5)		&	N/A	\\
\multicolumn{1}{c|}{ {\bf $^6$Li $1^+_{\rm g.s.}$} }  	&		&		{\bf NCSM}$^{\rm b}$	&				{\bf VMC [GFMC]}$^{**}$	&		\\
		BE (MeV)&	30.45		&	30.95		&	27.0(1) [$31.2(1)$]$^{\rm c}$	&	31.99	\\
		$r_{\rm p}$ (fm) 	&	2.11		&	2.13		&	2.46(2)$^{\rm d}$ 		&	2.43$^{\rm e}$  	\\
			&	2.13$^{***}$	 	&	2.14$^{***}$		&			&		\\
		$Q$ ($e$ fm$^2$)	&	$-0.080$		&	$-0.064$		&	$-0.33(18)^{\rm d}$		&	$-0.0818(17)$	\\
		$\mu$ ($\mu_N$)	&	+0.839		&	+0.838		&	+0.828(1)$^{\rm d}$		&	+0.82205	\\
 \multicolumn{1}{c|}{{\bf $^{12}$C}	} 	&					&	{\bf NCSM}$^{\rm f}$				&			&				\\
   		BE (MeV)	&		85.95			&		87.90			&			&		92.16		\\
		$E_{2^+_1}$ (MeV)		&		4.64			&		4.69			&			&		4.44		\\
		       $Q_{2^+_1}$  ($e$ fm$^2$)		&	    	 +3.735	 		&	      +3.741    			&			&	   +6(3)		\\
		       $\mu_{1^+_1}$ ($\mu_{\rm N}$)  	&	0.839		&	0.848		&			&	   N/A      	\\
		       $B(M1; 1_1^+ \rightarrow 0^+_{\rm g.s.})$ ($\mu_{\rm N}^2$)    	&	0.012		&	0.013		&			&	  0.0145(21) 		\\
\end{tabular}
\end{center}
\begin{tabnote}
$^{*}$N3LO-EM NN; $^{**}$AV18/Urbana IX; $^{***}$NNLO$_{\rm opt}$; $^{\rm a}$From Ref. \cite{BakerLBND20}; $^{\rm b}$From Ref. \cite{DytrychLMCDVL_PRL12};  $^{\rm c}$From Ref. \cite{WiringaS98};  $^{\rm d}$From Ref. \cite{PudlinerPCPW97}, without contributions from two-body currents;   $^{\rm f}$From Ref. \cite{DytrychMLDVCLCS11};  $^{\rm e}$Deduced from the $^{6}$Li charge radius of 2.56(5) fm \cite{LiSWY71}.
\end{tabnote}
\end{table}

Similarly, for the $^6$Li ground state and low-lying isospin-zero states, Ref. \cite{DytrychLMCDVL_PRL12}  has validated the use of selected SA spaces as compared with the complete $N_{\rm max}=12$ model space, as  illustrated in Table \ref{benchmark} for selected observables for 14 shells, \hw=20 MeV, and with JISP16 and NNLO$_{\rm opt}$ $NN$ interactions.
In Ref. \cite{DytrychHLDMVLO14}, these results are compared with those of the {\it ab initio} VMC and GFMC methods using the AV18 NN and Urbana IX 3N  interactions (see \citenum{RevModPhys.87.1067,LynnTGL19}). We note the remarkable agreement, despite the use of  realistic interactions different in construction and properties (e.g., non-local vs. local). The close agreement between the the SA-NCSM and VMC results holds also for the $^6$Li point-proton density (Fig. \ref{enspectra}a), where the SA-NCSM calculations span model spaces of 14 shells ($N_{\rm max}=12$) that include all symplectic irreps (for $\hw=20$ MeV) or only the single symplectic irrep,   used to determine the corresponding $^6$Li energies and $B(E2)$ strengths shown in Fig. \ref{enspectra}a.

Results for a heavier nucleus, $^{12}$C, corroborates the findings for $^{4}$He and $^{6}$Li~\cite{DytrychMLDVCLCS11,SargsyanLBDD20}. Selected SA-NCSM observables are listed in Table \ref{benchmark} for 10 shells and \hw=20 MeV, and practically coincide with the complete-space calculations. In addition, Ref. \cite{DytrychMLDVCLCS11} has shown that the size of the model space and the number of nonzero Hamiltonian matrix elements -- for SA selected spaces -- grow slowly with the model-space size $N_{\rm max}$. 

\begin{figure}[h]
\includegraphics[width=\textwidth]{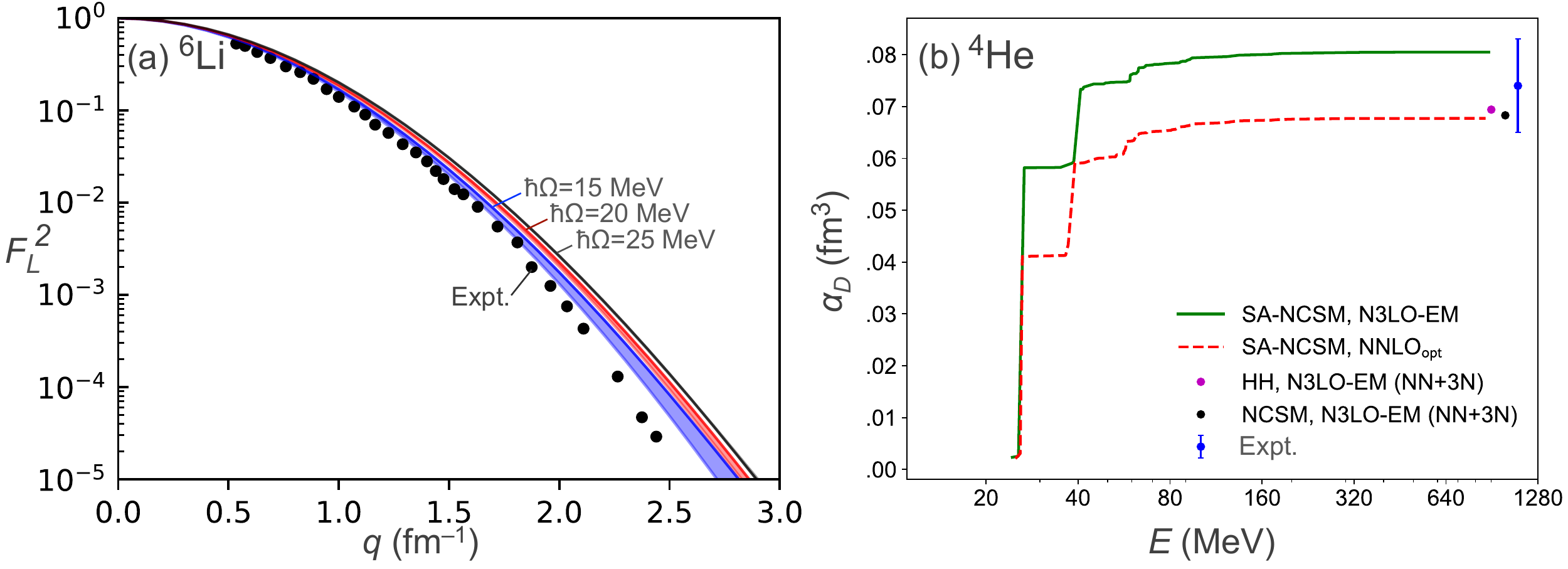}
\caption{
(a) Longitudinal  electric charge form factor $F_L^2$ (translationally invariant and adjusted to account for the finite proton size) for the $^6$Li ground state  
(based on {\it ab initio} SA-NCSM calculations  with NNLO$_{\rm opt}$ in various selected model spaces up through 14 shells).
Electron scattering experimental data are taken from Ref. \cite{LiSWY71}.
Data from Dytrych et al. (2015) \cite{DytrychHLDMVLO14}.
(b) Electric dipole polarizability $\alpha_D$ (translationally invariant) as a function of the excitation energy $E$ (using N3LO-EM and NNLO$_{\rm opt}$ chiral potentials, and {\it ab initio} HH calculations, along with  SA-NCSM and NCSM calculations in a model space of 18 HO shells with \hw=25 MeV inter-shell distance).  Figure adapted with permission from Baker et al. (2020); copyright 2020 \cite{BakerLBND20}.
}
\label{fig:alphaD}
\end{figure}

Furthermore, the SA framework has been applied to observables that can be extracted from  electron scattering and photoabsorption experiments. Ref. \cite{DytrychHLDMVLO14} has studied the longitudinal  electric charge  form factor   using  \textit{ab initio} SA-NCSM calculations for the $1^{+}$ ground  state of $^{6}$Li, as shown in Fig. \ref{fig:alphaD}a, where  \SU{3}-selected spaces in 14 shells  (light-colored bands) are compared with the corresponding complete model space (solid lines).  The agreement, first with NCSM and also with experiment, points to the fact that the symmetry considerations of the type we consider in the SA framework properly treat, in addition, 
 excitations to higher HO shells relevant for typical momentum transfers, $q \lesssim 4$ fm$^{-1}$  \cite{DytrychHLDMVLO14}.

The electric dipole polarizability $\alpha_D$ can be extracted from photoabsorption experiments. Specifically, $\alpha_D $ can be deduced from the  photoabsorption cross sections $\sigma_{\gamma}(E)=4\pi^2 \alpha E R(E)$  by integrating the data~\cite{Arkatov:1974, Arkatov:1980} with the proper energy weight, where $R(E)$ is the  dipole response function  for given excitation energy $E$ and $\alpha$ is the fine-structure constant. The inverse energy weighted sum rule for $^4$He  can be used to calculate $\alpha_D$, based on the relation $\alpha_D = 2 \alpha \times {\rm  IEWSR}$,  which is compared  with experiment (Fig.~\ref{fig:alphaD}b).
 In particular, Ref. \cite{BakerLBND20} has shown that  the N3LO-EM yields a larger $\alpha_D$ value as compared with the NNLO$_{\rm opt}$, while both results fall within the experimental uncertainties. This is consistent with earlier theoretical work,  which included the complementary 3N forces in the N3LO-EM and has shown that the 3N forces reduce the value of $\alpha_D$ by as much as 15\% \cite{Gazit_PRC_2006}. A remarkable result is that the outcome for the  N3LO-EM (NN$+$3N), calculated in the HH \cite{Ji_PRL_2013} and the NCSM \cite{PhysRevC.79.064001},  closely agrees with that for the NNLO$_{\rm opt}$ using only NN forces. 

Within the SA framework, calculations are feasible up through medium-mass nuclei. E.g., first no-core shell-model calculations in 10 shells  are now available for $^{48}$Ti  (Fig \ref{enspectra}c)  including the ground-state one-body density profile (in the \textit{body-fixed} frame) and an estimate for the quadrupole moment of its lowest $2^+$ state that is in a good agreement with experiment. These calculations use SA model spaces with about $6\times10^{6}$ basis states, compared with the unfeasible dimension of $3\times10^{13}$ of the corresponding complete model space. 

As another illustrative example, we show structure observables for  $^{20}$Ne, together with its ground-state one-body density  (Fig \ref{enspectra}b) and response to an isoscalar electric monopole  probe $M_0={1 \over 2} \sum_i r_i^2$ (Fig. \ref{ff_lsr}b). In Fig. \ref{ff_lsr}b, this is illustrated by the Lorentz integral transform \cite{Efros:1994_PLB,bacca:2013_prl} for monopole transitions to the $^{20}$Ne ground state using a Lorentzian kernel width of $\Gamma_{\mathfrak{L}}=2$ MeV that yields the  response function  in the $\Gamma_{\mathfrak{L}} \rightarrow 0$ limit. The first large peak is associated with  a breathing mode, or giant monopole resonance \cite{BakerThesis19}, and can, in turn, provide a stringent probe on incompressibility and nuclear saturation properties \cite{GARG201855}.
Indeed, since  the $M_0$ operator is a symplectic generator and does not mix symplectic irreps, the monopole response of Fig. \ref{ff_lsr}b tracks the contribution of the  predominant shape of  the $^{20}$Ne ground state to all excited $0^+$ states. It is not surprising then that the distribution and the peak of the response  are consistent with the results of  Ref. \cite{DytrychLDRWRBB20} (see the higher $0^+$ states in Fig. \ref{Sp_pict}a). Indeed, the set of excited $0^+$ states in Fig. \ref{Sp_pict}a with nonnegligible contribution of the \ph{1} excitations of the ground-state equilibrium shape has been suggested in Ref.  \cite{DytrychLDRWRBB20} to describe a fragmented giant monopole resonance with a centroid around $29$ MeV  and a typical wave function spread out to higher deformation due to vibrations \cite{BahriR00}, in contrast to the ground state. This is clearly evident in the $\beta$-$\gamma$ plots in Fig. \ref{Sp_pict} that depict the deformation distribution within the same symplectic irrep for the ground state and the GR peak across the average deformation $\beta$ and triaxiality $\gamma$.

\section{NUCLEAR REACTIONS WITH SYMMETRY-ADAPTED BASIS}
\label{SAreactions}

\subsection{Alpha-Induced Reactions} 
\label{alpha}
Partial widths are given by the decay rates of resonances into different open channels. They are not directly measurable, and extraction is  model-dependent  to a greater or lesser extent.  Alpha widths  and alpha capture reactions of intermediate-mass nuclei  are now feasible in the \textit{ab initio} SA framework, including intermediate-mass nuclei along the path of x-ray burst nucleosynthesis.  In general, the formalism is applicable up through the medium-mass region  with the \textit{ab initio} SA-NCSM, and for heavier nuclei, e.g., when nuclear fragments are described in the no-core symplectic shell model (NCSpM) with effective many-nucleon interaction  \cite{DreyfussLTDB13,TobinFLDDB14,DreyfussLTDBDB16,BahriR00}. The NCSpM can reach ultra-large model spaces, and has achieved successful no-core shell-model descriptions of low-lying states in deformed $A=8$-$24$ nuclei \cite{TobinFLDDB14}, and in particular, of the  elusive Hoyle state in $^{12}$C and its first $2^+$ and $4^+$ excitations \cite{DreyfussLTDB13}. 

Modeling  nuclear systems with cluster substructure represents a major challenge for many-particle approaches that build on realistic interactions. For light nuclei, there  has been recent progress in $\textit{ab initio}$ descriptions of alpha cluster systems, including the
Green's function Monte Carlo method with applications to the $\alpha$-cluster structure of  $^{8}$Be and  $^{12}$C, along with electromagnetic transitions \cite{RevModPhys.87.1067}; the nuclear lattice effective field theory with applications to the Hoyle state energy and the astrophysically relevant $\alpha$-$\alpha$ scattering problem \cite{EpelbaumKLM11,PhysRevLett.111.032502,ElhatisariLRE15}; and the hyperspherical harmonics method, with applications to giant resonance modes in  $^{4}${He} \cite{BaccaBLO13}. Of particular note are recent developments that combine RGM with configuration-interaction methods \cite{KravvarisV19,mercennemp19}, as well as with $\textit{ab initio}$ no-core shell model and SA-NCSM \cite{PhysRevLett.101.092501,BarrettNV13,MercenneLEDP19} (cf. Sec. \ref{sargm}).
For a review of cluster models, see Ref. \cite{FreerHKLM18},  including some of the earliest techniques that treat particles within localized clusters, such as RGM \cite{Wheeler_PR52_1937a, WildermuthT77} and the related generator coordinate method  \cite{Horiuchi_PTP43_1970}, as well as molecular dynamics approaches \cite{KanadaEnyo98,ChernykhFNNR07}.

Ref. \cite{DreyfussLESBDD20} has recently presented a new many-body technique for determining challenging alpha widths and asymptotic normalization coefficients (ANCs) utilizing  \textit{ab initio} SA-NCSM  wave functions, with a focus on the $^{16}$O($\alpha$,$\gamma$)$^{20}$Ne reaction rate. Indeed, the SA framework is ideal for addressing cluster substructures,  as it enables large model spaces needed for clustering,  and  capitalizes on the complementary nature of the symplectic basis and the cluster basis \cite{HechtZ79,Suzuki86,SuzukiH86}.  
Several studies have taken advantage of this relationship using a single \SU{3} deformation for the clusters. In particular, this approach has been used to describe the sub-Coulomb $^{12}$C$+ ^{12}$C resonances of  $^{24}$Mg \cite{SuzukiH82} of particular interest in astrophysics, as well as  spectroscopic factors for alpha conjugate nuclei (that is, nuclei with multiples of two protons and two neutrons) \cite{Suzuki86,SuzukiH86,HechtRSZ81}.
These studies have shown that some of the most important shell-model configurations can be expressed by exciting the relative-motion degree of freedom of the clusters. Further, they have indicated that an approach that utilizes both the cluster and symplectic bases proves to be advantageous, especially since the model based on the cluster basis only, for clusters without excitations, tends to overestimate cluster decay widths and underestimates $E2$ transition rates \cite{SuzukiH86}.

\begin{figure}[h]
\includegraphics[width=0.7\textwidth]{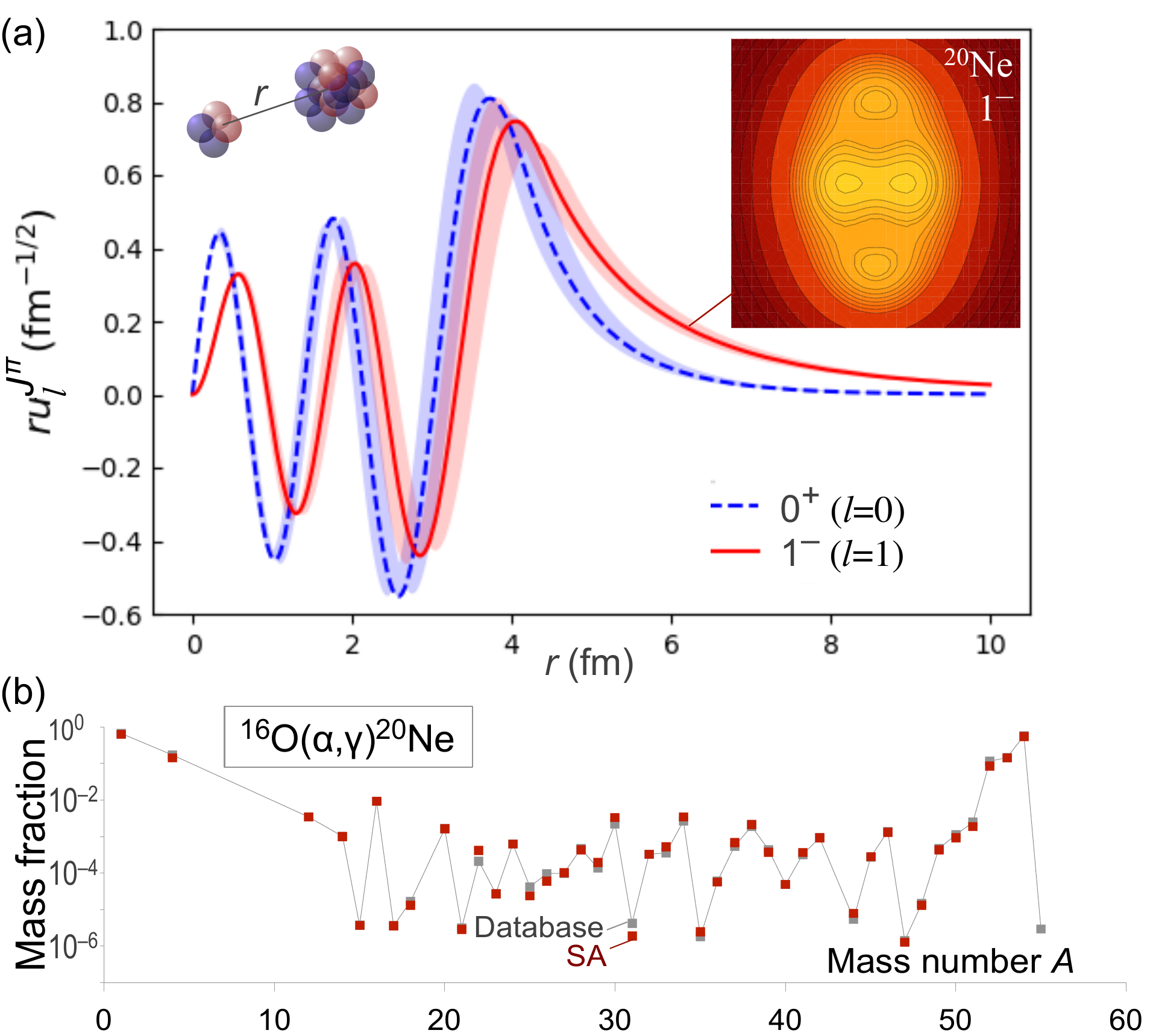}
\caption{(a) 
$\alpha+^{16}$O $l=0$ and $l=1$ relative wave functions  from \textit{ab initio} SA-NCSM calculations of the $^{20}$Ne ground state and lowest $1^-$ state, respectively (with NNLO$_{\rm opt}$  and  \hw=13-17 MeV inter-shell distance). (b)  Difference between the initial mass fractions of the neutron star and the mass fractions 24 hours after the burst begins based on the MESA XRB simulation, showing good agreement between the pattern from the SA-calculated reaction rate and the pattern from the database reaction rate. All isotopes in the network with mass differences greater than $10^{-10}$ are shown.
Figure adapted with permission from Dreyfuss et al. (2020); copyright 2020 \cite{DreyfussLESBDD20}.}
\label{alpha_reac}
\end{figure}

In  Ref. \cite{DreyfussLESBDD20}, the first alpha partial width of the lowest $1^-$ resonance has been reported based on  \textit{ab initio}  $^{20}$Ne wave functions.
Specifically, for the partition into $a$- and $A$-particle clusters, the relative wave function $ru_{c l}^{J^{\pi}}(r)$  is given by
\begin{equation}
    \label{eq:SAwithOverlap}
    u_{c l}^{J^{\pi}}(r)
      =
    \sum_{\eta } R_{\eta l}(r)
    \langle
    (A+a)\mathfrak{a} J^{\pi}M | ((A) \mathfrak{a}_1I_1^{\pi_1}, (a) \mathfrak{a}_2I_2^{\pi_2})I,\eta l;J^{\pi}M\rangle,
\end{equation}
where the cluster system is defined for a channel $c  = \{ \mathfrak{a}, \mathfrak{a}_1, I_1^{\pi_1}, \mathfrak{a}_2, I_2^{\pi_2},I\}$, which is labeled by the angular momentum  (spin) and parity of each of the clusters  and the total spin of the clusters $I$ (the labels $\mathfrak{a}$, $\mathfrak{a}_1$ and $\mathfrak{a}_2$ denote all other quantum numbers needed to fully characterize their respective states), and a partial wave $l$. $R_{\eta l}(r)$ is the  single-particle HO radial wave function ($\eta=0,1,2,\dots$ label  $s$, $p$, $sd$,\dots major HO shells). The integral of Eq. (\ref{eq:SAwithOverlap}) over $r$ yields a spectroscopic factor. The  overlap $ \langle
    (A+a)\mathfrak{a} J^{\pi}M | ((A) \mathfrak{a}_1I_1^{\pi_1}, (a) \mathfrak{a}_2I_2^{\pi_2})I,\eta l;J^{\pi}M\rangle$ is calculated for the $(A+a)$-body state of the composite system (in the present example, $^{20}$Ne) and the cluster configurations (in the present example, $\alpha$+$^{16}$O), using an efficacious \SpR{3} group-theoretical technique \cite{Suzuki86,SuzukiH86,DreyfussLESBDD20}. In addition, based on the microscopic ${ R }$-matrix approach \cite{DescouvemontB10}, $ru_{c l}^{J^{\pi}}(r)$ is  matched at a channel radius to the exact solution to the Coulomb potential in the exterior (shown in Fig. \ref{alpha_reac} at large distances). In doing this, one is able to obtain the two-cluster wave function that reflects the microscopic structure of the fragments while having the correct asymptotics, and hence calculate $\alpha$ widths for resonances and  asymptotic normalization coefficients (ANC's) for bound states \cite{DreyfussLESBDD20}.

The new method is applied to the $1^{-}$ resonance in $^{20}$Ne with a known natural width,  and because the state decays entirely through $\alpha$ emission, the natural width is the $\alpha$ partial width. 
Specifically,  the $\alpha+^{16}$O $l=0$ and $l=1$ wave functions are calculated using  the \textit{ab initio} SA-NCSM for the $^{20}$Ne ground state and lowest $1^-$ state in 11 shells  (Fig. \ref{alpha_reac}, where bands are given by \hw=13, 15, 17 MeV).
Using extrapolations that do not dependent on the channel radius, Ref. \cite{DreyfussLESBDD20} reports a value of $\Gamma_{\alpha}=10(3)$ eV for the alpha partial width of the $1^{-}$ resonance, with uncertainty given by the variation in $\hbar\Omega$. Given that no parameters are fitted to nuclear data in this study, this estimate agrees reasonably well with the natural width of the $^{20}$Ne $1^{-}$ state of $28(3)$ eV \cite{PhysRevC.22.356,Constantini_PRC82_2010}. We note that while experimental thresholds are used in Ref. \cite{DreyfussLESBDD20}, the study has emphasized the key role of  correlations in developing cluster structures and collective modes, without which  widths become drastically reduced.

This method also allows for first estimates for ANCs in $^{20}$Ne within a no-core shell-model framework. The extrapolated ANC for the ground state is estimated to be $C_{0}=3.4\pm1.2\times10^{3}\,\mathrm{fm}^{-1/2}$ from \textit{ab initio} SA-NCSM calculations.  For the first excited  $4^{+}$ state in  $^{20}$Ne  that lies in close proximity to the $\alpha+^{16}$O threshold, the ANC is estimated from $N_{\rm max}=14$ NCSpM calculations to be an order of magnitude larger \cite{DreyfussLESBDD20}.

The alpha widths can, in turn,  be used to calculate alpha capture reaction rates for narrow resonances of interest to astrophysics. This is achieved by using the narrow resonance approximation, for which reaction rates are given by
\begin{equation}
    \label{eq:narrowres}
    N_{A}\langle\sigma v\rangle_{r}
    =
    \frac{1.539\times10^{11}}{(\mu_{A,a} T_{9})^{3/2}}
    e^{-11.605E_{r}/T_{9}}
    (\omega\gamma)_{r},
\end{equation}
where $T_{9}$ is temperature in GK, $\mu_{A,a} $ is the reduced  mass of the two clusters, $E_{r}$ is the resonance energy  in MeV, and the resonance strength is defined as
\begin{equation}
    (\omega\gamma)_{r}
    =
    \frac{2J+1}{(2I_{1}+1)(2I_{2}+1)}
    \frac{\Gamma_{\alpha}\Gamma_{\gamma}}{\Gamma}.
\end{equation}
Using the SA estimate for the alpha width $\Gamma_{\alpha}$ and   $\Gamma_{\gamma}/\Gamma$ extracted from the resonance strength of 
Ref. \cite{Constantini_PRC82_2010}, the contribution to the $^{16}$O$(\alpha,\gamma)^{20}$Ne  reaction rate through the 1.06-MeV $1^{-}$ resonance in $^{20}$Ne is calculated at astrophysically relevant temperatures. This calculated reaction rate is used as input to the Modules for Experiments in Stellar Astrophysics (MESA) code suite \cite{Paxton2019} to determine its impact on the abundance pattern produced during an x-ray burst event (Fig. \ref{alpha_reac}b). The MESA release \cite{Paxton2015} includes a model for an XRB with a constant accretion rate and consistent burning across the entire surface of the neutron star, based on GS 1826-24, also known as the ``clocked burster'' \cite{Ubertini_APJ514_1999}. This model is designed for a nuclear network of 305 isotopes, including proton rich isotopes up to $^{107}${Te}, but is also stable for a nuclear network of 153 isotopes up to $^{56}${Fe}, used in the present calculations. MESA includes all known reactions involving these nuclei, with reaction data taken from the REACLIB database \cite{Cyburt10}. Remarkably, the SA calculated reaction rate for the alpha capture reaction $^{16}$O$(\alpha,\gamma)^{20}$Ne is found to produce practically the same XRB abundance pattern as the known reaction rate available in the REACLIB database, as shown in Fig. \ref{alpha_reac}b.

\subsection{Scattering and  Reactions for a Single-Nucleon Projectile}
\label{sp} 
The section  presents a novel \textit{ab initio} symmetry-adapted framework for reactions based on the RGM  \cite{Mercenne:2019LDEP,MercenneLEDP19}, applicable to  nucleon scattering and capture reactions  with  light to medium-mass nuclei at the astrophysically relevant energy regime 
(Sed. \ref{sargm}). As illustrative examples, we discuss results for neutron scattering off $^{16}$O  and $^{20}$Ne \cite{LauneySOTANCP42018}. 
In addition, this section discusses  a state-of-the-art few-body approach to scattering at intermediate energies based on the multiple scattering theory \cite{Elster:1996xh,Dussan:2014vta,BurrowsBEWLMP20}, with a focus on \textit{ab initio}  scattering cross sections and spin reaction observables at energies $\sim 100-200$ MeV (Sec. \ref{multiscatt}), 

An important outcome of the RGM and multiple scattering approaches is an \textit{ab initio} nucleon-nucleus effective potential.
An alternative approach  has employed the Green's function framework and applied to low energies ($\lesssim20$ MeV per nucleon) using the self-consistent Green's function method  \cite{idini19} and coupled-cluster method \cite{RotureauDHNP17} (see also \citenum{FRIBTAwhite2018}).
These studies have built upon earlier theoretical frameworks, such as the one  introduced by Feshbach, leading to the Green's function formulation \cite{CapMah:00}, and the one pioneered by Watson~\cite{Watson1953a,KMT} for elastic scattering of a nucleon from a nucleus, leading to the spectator expansion of the multiple scattering theory~\cite{Siciliano:1977zz}.  Indeed, recent progress has been made to derive microscopic optical potentials, 
which, in turn, can be used to  provide cross sections for elastic scattering, as well as input to (d,p) and (d,n) reactions \cite{Rotureau_2020}. These studies have emphasized the need for realistic interactions that correctly reproduce rms radii, as well as the importance of collective degrees of freedom to properly account for absorption.

\subsubsection{Low energies: resonating group method}
\label{sargm}

 The resonating-group method  \cite{WildermuthT77} is a microscopic method which uses fully antisymmetric wave functions, treats correctly the center-of-mass motion of the clusters, and takes  internal correlations of the clusters into consideration. 
 In the RGM, nucleons are organized within different groups, or clusters, ``resonating'' through the inter-cluster exchange of nucleons. The antisymmetrization between the different clusters enforces the Pauli exclusion principle. All of these features make this method particularly suitable for providing unified descriptions of nuclear structure and reaction observables. It builds upon the successful combination of the RGM and NCSM with NN and 3N interactions  for light nuclei \cite{QuaglioniN09}. 
With the use of the SA basis, the SA-RGM  expands \textit{ab initio} reaction theory to reactions of heavier nuclei and weakly bound systems near the drip lines for astrophysically relevant energies.

Traditionally, RGM adopts  generalized cluster wave functions as basis functions, which describe the motion of a system of two or more clusters. 
We consider two nuclear fragments, or two-cluster nuclear reactions.
For two clusters $A$ and $a$, the cluster states for a channel $c$ are defined as (cf. Sec. \ref{alpha}):
  \begin{equation}{ \ket{ { \Phi }_{ c r }^{J^\pi} }= { \{ \{{ \ket{ (A) \mathfrak{a}_1 I_1^{\pi_1}} \times \ket{ (a) \mathfrak{a}_2 I_2^{\pi_2}} } \}^I \times Y_{ \ell } ({ \hat{ r } }_{ A,a }) \}^{J^\pi} } \frac{ \delta(r - { r }_{ A,a }) }{ r { r }_{ A,a } } }
    \end{equation}
for a relative distance between the clusters ${ r }_{ A,a }$ (cf. Eq. \ref{eq:SAwithOverlap}). The A+a nuclear wave function is given in terms of the cluster states 
  \begin{equation}
    \ket{ { \Psi } ^{J^\pi}} = \sum_{c} \int_{r} dr { r }^{ 2 } \frac{ g^{J^\pi} _{ c }(r) }{ r } { \mathcal{A}_c } \ket{ { \Phi }_{ c r }^{J^\pi} } \;,
    \label{RGM_ansatz}
  \end{equation}
with unknown amplitudes ${ { g }_{ c }^{J^\pi}(r) }$ that are determined  by solving the integral 
Hill-Wheeler equations (that follow from the Schr\"odinger equation): 
  \begin{equation}
    \sum_{c} \int dr { r }^{ 2 } \left[ { H }_{ c' c } (r',r) - E { N }_{ c'c }(r',r) \right] \frac{ { g }_{ c }^{J^\pi}(r) }{ r } = 0.
    \label{RGM_equations}
  \end{equation}
  Here,  $H _{ c'c }(r',r) ={ \bra{ { \Phi }_{ c' r' } ^{J^\pi}} { \mathcal{A} }_{c'} H { \mathcal{A} }_{c} \ket{ { \Phi }_{ c r }^{J^\pi} } }$ is the Hamiltonian kernel and $N_{ c' c }(r',r)={ \bra{ { \Phi }_{ c' r' }^{J^\pi} } { \mathcal{A} } _{c'} { \mathcal{A} }_c \ket{ { \Phi }_{ c r }^{J^\pi} } }$ is the norm kernel, where ${ { \mathcal{A} } }$ is the antisymmetrizer. The kernels are computed using the microscopic wave functions of the clusters that can be obtained in the \textit{ab initio} NCSM and SA-NCSM. Once the kernels are computed, Eq.(\ref{RGM_equations}) can then be solved using the microscopic ${ R }$-matrix approach \cite{DescouvemontB10}.
 
In the SA-RGM,  the target nucleus is described by SA-NCSM many-body wave functions. Specifically, a target state with spin and parity ${ I_1 }^{\pi_1}$ with projection ${ M_1 }$  is constructed in terms of the SA basis:
  \begin{equation}
    \ket{ (A) \mathfrak{a}_1 I_1^{\pi_1} M_1 } = 
   \sum_{ { \mathfrak{b} }_1  \omega_1\kappa_1L_1S_1} { C }_{ { \mathfrak{b} }_1 }^{ { \omega }_{ 1 } { \kappa }_{ 1 } L_1{ S }_{ 1 } }  \ket{ { \mathfrak{b} }_1  \omega_1 \kappa_1 (L_1S_1)  I_1^{\pi_1} M_1 },
    \label{SU3wf}
  \end{equation}
  where the labels are defined, in general, as 
  $\mathfrak{b} \equiv  \left\{ \dots   {\omega}_{ \rm p } { \omega}_{\rm n } \rho N; { S }_{ \rm p } { S }_{\rm n } \right\}$  
  and deformation $\omega \equiv (\lambda\, \mu) $. Protons and neutrons are labeled by p and n, respectively,  and $S$ labels the intrinsic spin (``$\dots$" denotes all additional quantum numbers). The \SU{3} outer multiplicity ${ \rho }$ \cite{DraayerLPL89} results from the coupling of the proton deformation with that of neutrons to total deformation $\omega_1 $. As mentioned above,   ${ N }$ labels the total HO excitations ($N\leq N_{\rm max}$).  For a single-particle projectile, the SA-RGM basis states can be thus defined for a channel $\{{\nu_1;\nu}\}=\{{ \omega }_{ 1 } { \kappa }_{ 1 } (L_1{ S }_{ 1 }); \omega \kappa (L S)\}$ as:
  \begin{equation}
    \ket{ { \Phi }_{ \nu_1; \eta }^{ \nu J^\pi M } } = \sum_{ { \mathfrak{b} }_{ 1 } } 
    { C }_{ { \mathfrak{b} }_{ 1 } }^{ \nu_1 } 
    { \left\{ \ket{ { \mathfrak{b} }_{ 1 } { \omega }_{ 1 } { S }_{ 1 } } \times \ket{ (\eta \, 0) {1 \over 2} } \right\} }^{ \nu J M },
    \label{SU3RGMstates}
  \end{equation}
  where the \SU{3} basis states for the target  are coupled  to the HO single-particle states ${ \ket{ (\eta \,0) {1 \over 2} } }$ of the projectile.
 We note that there is no dependence on the orbital momentum of the projectile, only on the shell number it occupies, $\eta$. Furthermore, the summation over 
 ${ { \mathfrak{b} }_{ 1} }$ 
 implies that the SA-RGM basis requires only a part of the information present in the SA basis. 
 
 The SA-RGM basis is used to calculate the RGM kernels, which is the main computational task in RGM  \cite{QuaglioniN09}. These include the norm kernel, which is the overlap between antisymmetrized non-orthogonal RGM basis states. It consists of a direct part (a Dirac delta functions), which dominates at large relative distances, and an exchange part that takes into account the Pauli principle at short distances. The exchange norm kernel is related to the permutation operator ${ { { P } }_{  } }$ that exchanges the nucleon projectile with another nucleon within the target \cite{QuaglioniN09}. The exchange norm kernel in the SA-RGM basis is thus reduced to evaluating the following (similarly, for the Hamiltonian kernels):
  \begin{eqnarray}
     \bra{ { \Phi }_{\nu_1' ; \eta'  }^{ \nu' JM } } { { P } }_{  } \ket{ { \Phi }_{ \nu_1; \eta }^{ \nu JM } } 
     &=& { \delta }_{ \nu' \nu } \sum_{ { \omega }_{ o } { S }_{ o } { \rho }_{ 0} } { \Pi }_{ { S }_{ o } { S }_{ 1 }' } { (-1) }^{ \eta + \eta' - { \omega }_{ o } } { (-1) }^{ { S }_{ 1 } + \frac{ 1 }{ 2 } + S' } 
     \WignerSIXj{ { S }_{ 1 } }{ { S }_{ o } }{ { S }_{ 1 }' }{ \frac{ 1 }{ 2 } }{ S }{ \frac{ 1 }{ 2 } } 
     \nonumber \\
   & \times &  \sqrt{ \frac{ {\rm dim} { (\omega }_{ o }) }{ {\rm dim}(\eta\,0) } }   
    U\left[  { \omega }_{ 1 } { \omega }_{ o }  \omega' (\eta\,'0) ; { \omega}_{ 1 }' { \rho }_{ 0} 1 (\eta\,0) 1 1 \right] 
    { {\rho} }_{ \eta \eta' }^{  { \rho }_{ 0} { \omega }_{ o } { S }_{ o }} \left( \nu_1' ; \nu_1 \right), 
    \label{ExchangeMatrixSU3}
  \end{eqnarray}
  where ${ U\left[  \dots  \right] }$ is the \SU{3} 6-$(\lambda\,\mu)$ recoupling coefficient \cite{DraayerSU3_1}, analogous to the SU(2) 6-$j$ symbol, $\dim(\lambda\,\mu)={1\over 2}(\lambda+1)(\mu+1)(\lambda+\mu+2)$, and where the \SU{3} one-body density matrix elements are  defined as:
\begin{equation}
   { {\rho} }_{ \eta  \eta' }^{  { \rho }_{ 0} { \omega }_{ o } { S }_{ o } } \left( \nu_1' ; \nu_1 \right) = \sum_{ { \mathfrak{b} }_{ 1 } { \mathfrak{b} }_{ 1 }' } { C }_{ { \mathfrak{b} }_{ 1 }' }^{ \nu_1' } { C }_{ { \mathfrak{b} }_{ 1 } }^{ \nu_1}  \langle { { \mathfrak{b} }_{ 1 }' { \omega }_{ 1 }' { S }_{ 1 }' } ||| { \{ { a }_{ (\eta \,0) \frac{ 1 }{ 2 } }^{ \dagger } \times { \tilde a }_{ {(0\, \eta')}\frac{ 1 }{ 2 } } \} }^{ { \omega }_{ o } { S }_{ o } } ||| { { { \mathfrak{b} }_{ 1 } { \omega }_{ 1 } { S }_{ 1 } } } \rangle_{ { \rho }_{ 0} }.
  \label{O_objects}
\end{equation}
The matrix elements of the $ \rho$ density can be quickly computed in the SA basis, utilizing an efficacious \SU{3}-enabled vector-matrix-vector  algorithm, and this can be done prior to the computation of the kernels. It is notable that, as a result of the Kronecker delta function in Eq. \ref{ExchangeMatrixSU3}, the exchange part of the norm kernel turns out to be block-diagonal in this basis.
The reason is that the operator ${ { { P } }_{  } }$ is an SU(3) scalar and spin scalar, and therefore  preserves deformation and spin.
\begin{figure}[h]
\includegraphics[width=0.9\textwidth]{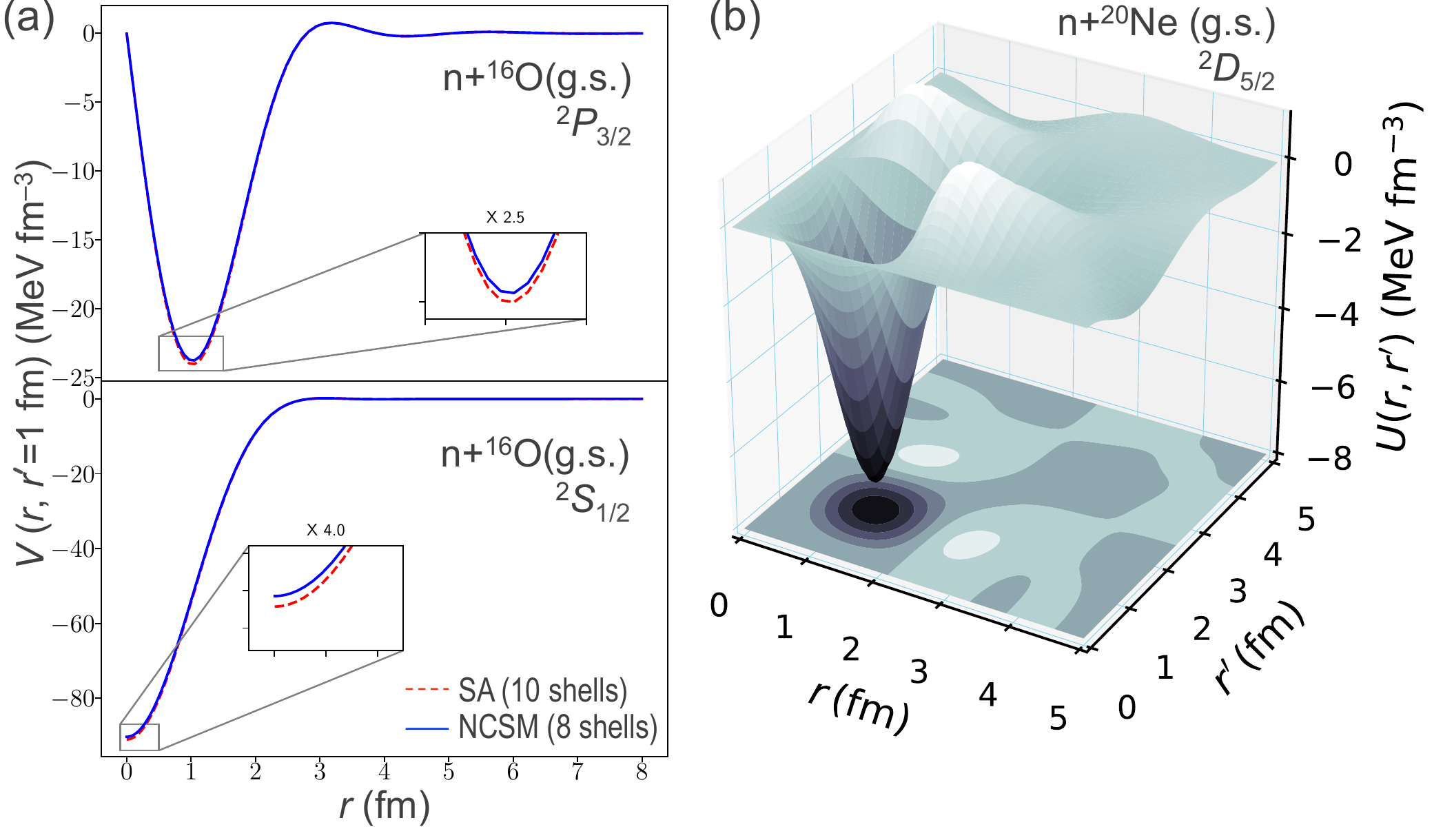}
\caption{(a) Translationally invariant potential kernel (direct, NN only) for n+$^{ 16 }$O(g.s) in the {\it ab initio} SA-RGM approach using  SA selected and complete model spaces
(based on {\it ab initio} SA-NCSM calculations  of $^{ 16 }$O  with \hw=16 MeV and NNLO$_{\rm sat}$ NN+3N in a model space of 10 shells; the projectile occupies 10 shells).  (b) Effective neutron-nucleus potential for the $^{ 20 }$Ne ground state, where effects of the target excitations and  antisymmetrization involving three nucleons are neglected (based on {\it ab initio} SA-NCSM calculations  of $^{ 20 }$Ne  with NNLO$_{\rm opt}$ in a model space of 11 shells and \hw=15 MeV inter-shell distance).
}
\label{sargm_pot}
\end{figure}

This procedure  allows the kernels to be calculated, for  each ${ J ^\pi M }$, through the SA-RGM channel basis of Eq. (\ref{SU3RGMstates}) that only depends on  the  deformation, rotation, and spin of the target $\nu_1$ (that is, ${ \omega }_{ 1 } { \kappa }_{ 1 }L_1 { S }_{ 1 }$), and the deformation, rotation, and spin of the target-projectile system $\nu$ (that is, $\omega \kappa LS $). Thus, the SA offers two main advantages: first, the number of unique \SU{3} configurations in the target wave function, we find, is a manageable number as compared with the complete model-space size, and second, a manageable number of  configurations for the target-projectile system is based on \SU{3} and SU(2) selection rules, namely, $\omega=\omega_1 \times (\eta \, 0)$ and $S=S_1 \times {1\over 2}$. Thus, for example, for proton or neutron scattering off $^{20}$Ne (with channels for ${ { 0 }^{ + } }$g.s.), there are only about $10^3$-$10^4$ SA-RGM basis  states for 7 to 13 shells, and only about $10^5$ for $^{23}$Mg when more target states are used (with channels for ${ 3/2_{\rm g.s.} ^{ + } }, { { 5/2 }^{ + } }$,${ { 7/2 }^{ + } }$). 
 Interestingly, the number of unique deformed configurations for heavier targets such as  Ne and Mg decrease in larger model spaces, as dominant shapes are allowed to develop, thereby reducing shape mixing.

As discussed above, it is important to validate the use of the SA basis and selected model spaces to ensure that the selection does not remove configurations relevant for these reaction processes. Indeed, a benchmark study for $^4$He and $^{16}$O has revealed that the selection has almost negligible effect on the norm kernels and potential kernels ${ \bra{ { \Phi }_{ c' r' } ^{J^\pi}} { \mathcal{A} }_{c'} V{ \mathcal{A} }_{c} \ket{ { \Phi }_{ c r }^{J^\pi} } }$ \cite{MercenneLEDP19,Mercenne:2019LDEP}, which are used as input to calculating  phase shifts and cross sections.  One such example is illustrated here for the direct potential kernel of  n+${ {  }^{ 16 } }$O(${ { 0 }^{ + }_{\rm g.s.} })$  (similarly, for a proton projectile)  with NNLO$_{\rm sat}$   up to 10 shells for  two partial waves ${ { S }_{ 1/2 } }$ and ${ { P }_{ 3/2 } }$ (Fig. \ref{sargm_pot}a). 
We note that, in these calculations, the 3N forces are included as a mass-dependent monopole interaction \cite{LauneyDD12}, which  has an effect on binding energies, and, for example, for the $^{16}$O ground-state energy,  
the  7-shell 3N contribution is 20.46 MeV, resulting in $-127.97$ MeV total energy for $N_{\rm max}=8$ and \hw=16 MeV, which agrees with the experimental value of $-127.62$ MeV.

In the SA-RGM framework, one starts from an \textit{ab initio} descriptions of all particles involved and derives the Hamiltonian kernel, which when orthogonalized, yields non-local effective nucleon-nucleus interactions for the channels under consideration.  For a single channel, if the effects of the target excitations are neglected, the non-local effective nucleon-nucleus interaction can be calculated for each partial wave, as illustrated for  n+${ {  }^{ 20 } }$Ne(${ { 0 }^{ + }_{\rm g.s.} }$) with NNLO$_{\rm opt}$ in 11 shells (Fig. \ref{sargm_pot}b). While these calculations limit the antisymmetrization to two nucleons only, this is a first step toward constructing effective nucleon-nucleus potentials for light and medium-mass nuclei  for the astrophysically relevant energies.

\subsubsection{Intermediate energies:  multiple scattering method}
\label{multiscatt}

To describe elastic scattering at intermediate energies, the \textit{ab initio} fully-consistent framework of the spectator expansion of the multiple scattering theory has been  developed at leading order \cite{BurrowsBEWLMP20}. It capitalizes on the  concept that  the two-body interaction between the projectile and the  nucleons inside the target nucleus play a dominant role. Hence, the leading-order term involves the interaction of the projectile and one of the target nucleons, the second-order term involves the projectile interacting with two target nucleons and so forth.  With the goal to derive an effective nucleon-nucleus potential, the effective potential operator is expanded  in terms of active particles ~\cite{Chinn:1993zza}. At leading order (two active particles), a consistent treatment requires an NN interaction to be used to calculate the NN transition amplitude (describing the interaction between the projectile and the struck target nucleon) as well as the microscopic structure of the target nucleus that enters by means of  one-body nuclear densities (for details, see \citenum{BurrowsBEWLMP20}). We note that the 3N effects enter only at the next order of the spectator expansion, and require two-body nuclear densities along with a solution to a three-body problem for three active nucleons.
\begin{figure}[th]
\includegraphics[width=\textwidth]{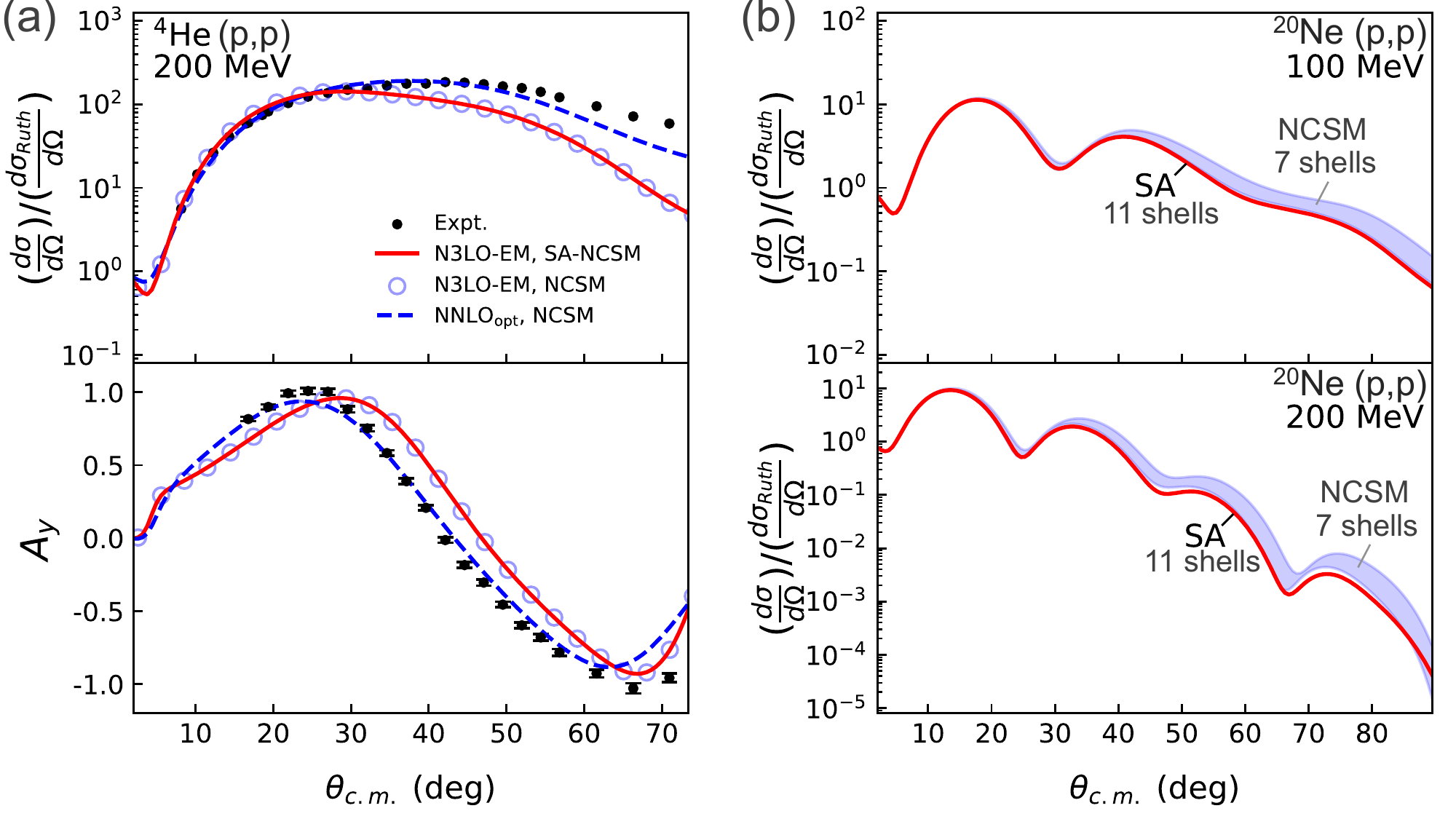}
\caption{
(a)  Angular distribution of the differential cross section divided by the Rutherford cross 
section and  analyzing power for elastic proton scattering on $^4$He at 200-MeV laboratory kinetic energy as a function of the center-of-mass (c.m.) angle,
showing perfect agreement between SA selected and complete (NCSM) model spaces of 15 shells ($N_{\rm max}=14$) with the N3LO-EM NN interaction and \hw=25 MeV. Also shown are calculations with the NNLO$_{\rm opt}$ NN chiral potential, in good agreement with the experiment.  
Experimental data from Ref.~\cite{Moss:1979aw}.
Figure adapted with permission from Burrows et al. (2020); copyright 2020 \cite{BurrowsBEWLMP20}.
(b) Angular distribution of the differential cross section divided by the Rutherford cross 
section for 100-MeV (top) and 200-MeV (bottom) proton laboratory kinetic energy on a $^{20}$Ne target (calculations use NNLO$_{\rm opt}$ with \hw=15 MeV inter-shell distance for SA and  \hw=14-16 MeV for the NCSM band).
}
\label{tmatrix}
\end{figure}

In a series of studies, the leading-order \textit{ab initio} effective nucleon-nucleus potential, which is nonlocal and energy-dependent, has been constructed \cite{Burrows:2017wqn,BurrowsEWLMNP19,BurrowsBEWLMP20}.
It has been used to calculate  reaction observables, such as cross sections and analyzing power $A_y$, in  He isotopes and other light nuclei, including $^{12}$C and $^{16}$O. For the first time,  the nuclear densities for the target in the multiple scattering theory have been derived from \textit{ab initio} calculations. The outcome of these studies reveals that the differential cross section and $A_y$ as a function of the center-of-mass angle, or equally the momentum transfer $q$, 
exhibit remarkable agreement with the experimental data when the chiral NNLO$_{\rm opt}$ NN potential is employed (see Fig.~\ref{tmatrix}a for $^{4}$He).

Similarly to reaction observables at low energies (Sec. \ref{sargm}), 
we show that the SA selected and complete model spaces practically coincide for  the angular distribution of the differential cross section and the analyzing power for protons on a $^4$He target  at 200~MeV laboratory projectile kinetic energy, using the N3LO-EM chiral potential  (Fig.~\ref{tmatrix}a).  Furthermore, the SA framework can extend calculations to intermediate-mass nuclei, namely, the \textit{ab initio} $^{20}$Ne(p,p)$^{20}$Ne differential cross section at 100 MeV and 200 MeV is studied and shown to exhibit a slight decrease as compared with smaller model spaces where the predominant shape is not fully developed (Fig.~\ref{tmatrix}b). Indeed, missing collective correlations have been suggested to reduce absorption in scattering at lower energies \cite{RotureauDHNP17}. The results in Fig.~\ref{tmatrix}b  pave the way toward exploring proton and neutron scattering on intermediate- and medium-mass targets, including  the role of collectivity and clustering.

\section{Summary and Outlook }

In summary, we discussed recent \textit{ab initio} developments made possible by the use of the SA basis that can reach ultra-large shell-model spaces in light up through medium-mass nuclei. 

\begin{summary}[SUMMARY POINTS]
\begin{itemize}
\item The SA basis exploits dominant symmetry in atomic nuclei, such as the symplectic \SpR{3} symmetry that does not mix nuclear shapes, and provides microscopic descriptions of nuclei in terms of collective shapes -- equilibrium shapes with their vibrations -- that rotate.  
\item Only a few shapes (a few symplectic irreps)  dominate in low-lying nuclear states, thereby making
significantly reduced SA selected model spaces ideal for study and prediction of various observables
for spherical and deformed open-shell nuclei.
\item  Small model spaces are sufficient to ``develop" many  shapes relevant to low-lying states, but often omit the vibrations of  largely deformed equilibrium shapes and spatially extended modes such as clustering; this is what is remedied by the use of selected model spaces. 
\item SA model spaces include all possible shapes, equivalently all particle-hole configurations, up to a given total particle-excitation energy and are selected only for larger energies; this implies that single-particle and collective degrees of freedom enter on an equal footing.
\item In the SA basis, the center-of-mass motion can be factored out exactly.
\item \textit{Ab initio} SA-NCSM calculations are now feasible for structure, reaction, and scattering observables of nuclei ranging from light to medium mass.
\end{itemize}
\end{summary}

The use of the SA basis is essential, first, for structure observables, especially for precise descriptions of cluster formations in nuclei (e.g., in $^{20}$Ne) or of collectivity in medium-mass nuclei, such as $^{48}$Ti of interest for neutrinoless double beta decay experiments that aim to determine whether the neutrino is its own antiparticle  \cite{PhysRevLett.124.232501,NovarioGEHJMNPQ20}. Second, the SA basis enables  couplings to the continuum, through excitations that are otherwise inaccessible and with the help of the SA-RGM basis, which accounts for decays to open channels. This is  critical for calculating reaction observables and for deriving nucleon-nucleus potential rooted in first principles, as discussed here in light of the SA-RGM approach for the astrophysically relevant energy regime and of the  multiple scattering method at intermediate energies. 
In many cases, results are highly sensitive to the microscopic structure, e.g., nucleon scattering and capture reactions at low energies are driven by a few open channels and isolated resonances, whereas collectivity and clustering is essential for alpha capture reactions and for deformed target or beam isotopes. 
As these approaches build upon first principles, they can probe features of the NN interaction that are relevant to reactions but remain unconstrained in fits to phase shifts or few-nucleon observables. 
 
 In short, with the help of high-performance computing resources,  the use of the SA concept in {\it ab initio} theory represents a powerful tool for the study of the structure and reactions of nuclei, and it is manageable as well as expandable; that is, one expects to be able to extend the reach of the SA  scheme from applications that are feasible today to the larger spaces and heavier nuclear systems of tomorrow, utilizing at each stage the predictive power of the {\it ab initio} approach to inform and support current and planned experiments.

\section*{ACKNOWLEDGMENTS}
We  acknowledge invaluable discussions with J. P. Draayer, S. Bacca, Ch. Elster, J. E. Escher, and S. Quaglioni, as well as  D. J. Rowe, J. L. Wood, G. Rosensteel, J. P. Vary, P. Maris, C. W. Johnson, and D. Langr. We also thank R. B. Baker, G. H. Sargsyan, A. C. Dreyfuss, and M. Burrows for providing important results.
This work was supported in part by the U.S. National Science Foundation  (OIA-1738287, PHY-1913728), SURA, the Czech Science Foundation (16-16772S), and the U.S. Department of Energy (DE-SC0019521). 
It benefitted from high performance computational resources provided by LSU (www.hpc.lsu.edu),  the National Energy Research Scientific Computing Center (NERSC), a U.S. Department of Energy Office of Science User Facility operated under Contract No. DE-AC02-05CH11231, as well as the Frontera computing project at the Texas Advanced Computing Center,  made possible by National Science Foundation award OAC-1818253.

\bibliographystyle{ar-style5}
\bibliography{launey_SA}

{\bf Related resources:}
\begin{itemize}
\item  D. J. Rowe and J. L. Wood, \textit{Fundamentals of nuclear models: foundational models} (World Scientific, Singapore, 2010).
\item I. J. Thompson and F. M. Nunes, \textit{Nuclear Reactions for Astrophysics} (Cambridge University Press, 2009).
\item  Y. Suzuki and R. G. Lovas and K. Yabana and K. Varga, \textit{ Structure and reactions of exotic nuclei} (Taylor \& Francis, London and New York, 2003).
\item V. K. B. Kota, \textit{{\rm SU(3)} Symmetry in Atomic Nuclei}, 
(Springer Singapore, 2020).
\item K. D. Launey, \textit{Emergent Phenomena in Atomic Nuclei from Large-Scale Modeling} (World Scientific Publishing Co., 2017).
\end{itemize}

\end{document}